\documentclass[%
 aip,
 amsmath,amssymb,
 reprint,%
]{revtex4-1}

\usepackage{graphicx}
\usepackage{dcolumn}
\usepackage{bm}

\usepackage[utf8]{inputenc}
\usepackage[T1]{fontenc}
\usepackage{mathptmx}
\usepackage{etoolbox}

\makeatletter
\def\@email#1#2{%
 \endgroup
 \patchcmd{\titleblock@produce}
  {\frontmatter@RRAPformat}
  {\frontmatter@RRAPformat{\produce@RRAP{*#1\href{mailto:#2}{#2}}}\frontmatter@RRAPformat}
  {}{}
}%
\makeatother
\begin{document}

\preprint{AIP/123-QED}

\title{Universality and non-differentiability: A new perspective on angular dispersion in optics}
\author{Layton A. Hall}
\affiliation{CREOL, The College of Optics \& Photonics, University of Central Florida, Orlando, FL 32816, USA}%
 \affiliation{Current address: Materials Physics and Applications - Quantum Division, Los Alamos National Laboratory, Los Alamos, NM 87545, USA}
 
\author{Murat Yessenov}%
 
\affiliation{CREOL, The College of Optics \& Photonics, University of Central Florida, Orlando, FL 32816, USA}%
\affiliation{Current address: Harvard John A. Paulson School of Engineering and Applied Sciences, Harvard University, Cambridge, MA, USA}

\author{Kenneth L. Schepler}
\affiliation{CREOL, The College of Optics \& Photonics, University of Central Florida, Orlando, FL 32816, USA}%

\author{Ayman F. Abouraddy}
\affiliation{CREOL, The College of Optics \& Photonics, University of Central Florida, Orlando, FL 32816, USA}%
\email{raddy@creol.ucf.edu}

\date{\today}

\begin{abstract}
Angular dispersion (AD) is a ubiquitous phenomenon in optics after light traverses a diffractive or dispersive device, whereby each wavelength propagates at a different angle. AD is useful in a variety of applications; for example, modifying the group velocity or group-velocity dispersion of pulsed lasers in free space or optical materials, which are essential ingredients in group-velocity matching and dispersion compensation. Conventional optical components introduce `differentiable' AD, so that the propagation angle can be expanded perturbatively around a fixed frequency, in which only a few low AD-orders are typically relevant. However, this model does not encompass newly emerging classes of propagation-invariant pulsed optical fields, such as `space-time wave packets', which incorporate a new form of AD that we call `non-differentiable AD'. This is a surprising feature: there exists a frequency at which the derivative of the propagation angle with respect to frequency is not defined. Consequently, the propagation angle cannot be expanded perturbatively at this frequency, and a large number of independently controllable AD orders are needed to approximate this condition. Synthesizing these new AD-induced field configurations requires constructing a `universal AD synthesizer' capable of accessing the magnitude and sign of any AD order, a capability missing from any single optical component to date. This Perspective article provides a unified schema for studying differentiable and non-differentiable AD, shows that non-differentiable AD enables circumventing many well-established constraints in optics -- thereby giving rise to new applications, and outlines the requirements for a universal AD synthesizer capable of producing both forms of AD.
\end{abstract}

\maketitle

\section{Introduction}\label{Section: Introduction_beginning}

\subsection{The role of angular dispersion (AD) in optics}

\begin{figure*}[t!]
\centering
\includegraphics[width=14cm]{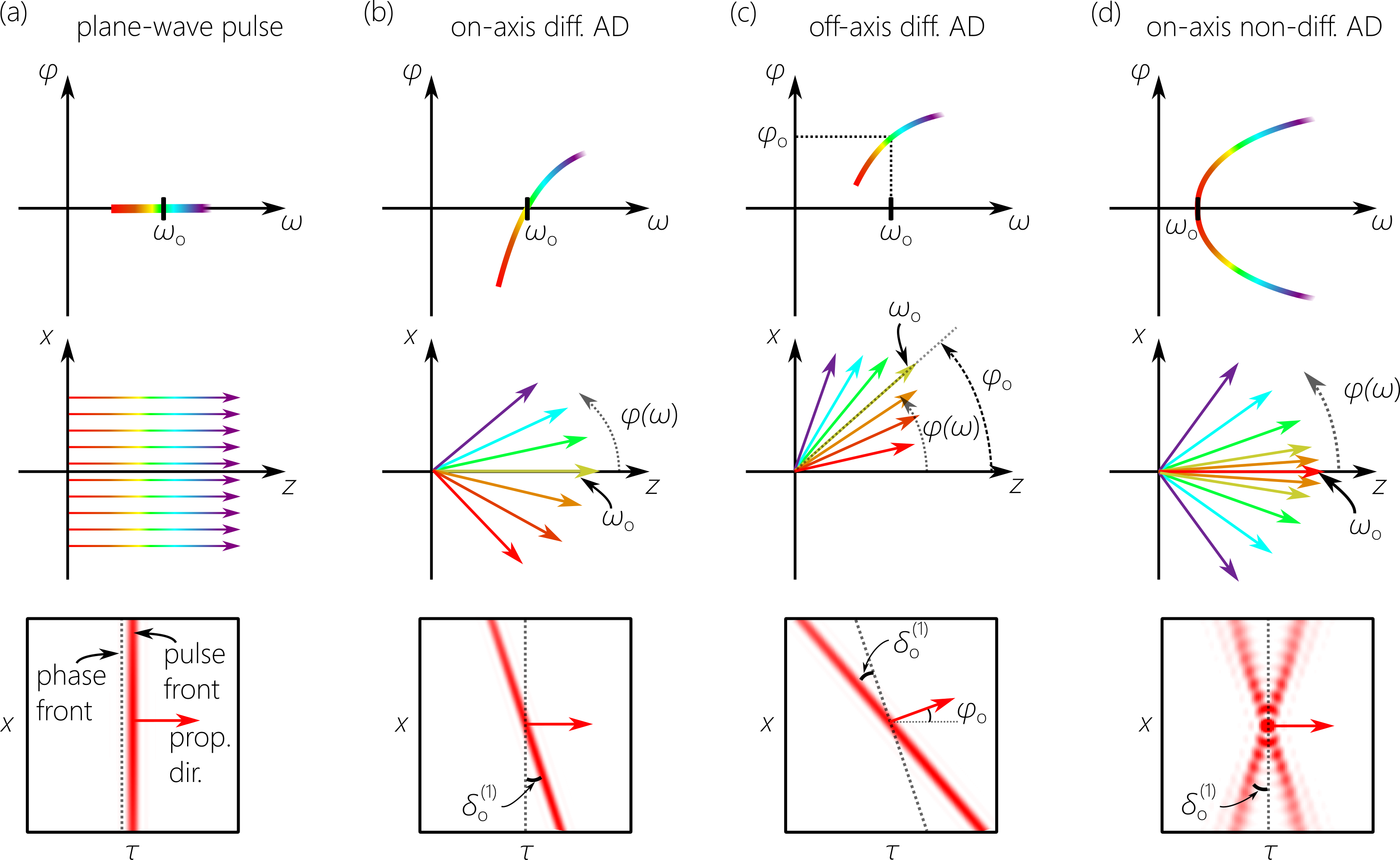}
\caption{Differentiable versus non-differentiable AD. The first row shows the AD profile $\varphi(\omega)$; the second the configuration in $(x,z)$-space; and the third the phase front (dotted lines) and the pulse front (solid curves) in $(x,t)$-space at fixed $z$. We consider the following field configurations: (a) a plane-wave pulse that is AD-free, $\varphi(\omega)=0$; (b) on-axis ($\varphi_{\mathrm{o}}=0$), differentiable AD; (c) off-axis ($\varphi_{\mathrm{o}}\neq0$), differentiable AD; and (d) on-axis, non-differentiable AD. The pulse front is tilted with respect to the phase front by an angle $\delta_{\mathrm{o}}^{(1)}$, defined in Eq.~\ref{Eq:AngleOfPulseFrontTilt}.}
\label{Fig:AngularDispersion}
\end{figure*}

A pivotal moment at the dawn of modern optics was Newton's observation of `angular dispersion', whereby wavelengths in initially collimated white light, upon traversing a prism, travel at different angles \cite{Sabra81Book}. Angular dispersion (AD) is a general optical phenomenon induced by dispersive or diffractive devices; e.g., prisms, diffraction gratings, or metasurfaces. Since the advent of the laser, AD has taken on an even more significant role in optics \cite{Torres10AOP,Fulop10Review}. Introducing AD into a collimated laser pulse tilts its pulse front (the plane of constant amplitude) with respect to its phase front (the plane of constant phase), which initially coincide [Fig.~\ref{Fig:AngularDispersion}(a)]. The result is called a `tilted pulse-front' (TPF) \cite{Bor93OE,Hebling96OQE} [Fig.~\ref{Fig:AngularDispersion}(b,c)]. Engineering the AD in a TPF gives rise to new phenomena that are utilized in ultrafast and nonlinear optics. For example, AD helps tune the TPF group velocity, which is utilized in group-velocity-matching scenarios, including THz generation via optical rectification \cite{Hebling02OE,Stepanov03APL,Hebling08JOSAB,Nugraha19OL,Wang20LPR}, and in nonlinear optical interactions, such as second-harmonic generation \cite{Volosov75Sov,Martinez89IEEE,Szabo90APB,Zhang90AO,Dubietis97OL,Richman98OL,Richman99AO,Schober07JOSAB}, parametric generators \cite{Szabo94APB,DiTrapani95PRA,Danielius96OL,Fulop07NJP,Major09RK,Isaienko09JOSAB,Piskarskas10IEEEJQE}, entangled-photon generation via spontaneous parametric downconversion \cite{Torres05PRA,Hendrych07OL,Shi08OL,Hendrych09PRA}, enhancing Raman conversion while suppressing self-phase modulation \cite{Klewitza98OC}, and spatiotemporal solitons \cite{DTrapanii98PRL,Liu00PRL,Liu00PRE,Wise02OPN}. Moreover, AD induces group-velocity dispersion (GVD) in free space \cite{Martinez84JOSAA,Fulop10Review,Torres10AOP}, which  is exploited in dispersion cancellation \cite{Treacy69IEEE,Fork84OL,Gordon84OL,Szatmari90AO,Szatmari96OL}. Other applications of TPFs include gain-matching via traveling-wave excitation \cite{Bor83APB,Klebniczki88APB,Hebling89OL,Hebling91JOSAB,Bleiner12AO} and dispersion compensation in chirped pulse amplification \cite{Pessot87OC}. New applications continue to emerge, including prolonged interaction between optical pulses and electron bunches for ultrafast electron diffraction \cite{Baum06PNAS,Pennacchio17SD}, laser-driven acceleration \cite{Plettner06PRSTAB,Peralta13Nature}, and X-ray generation \cite{Chang13PRL,Plettner08PTSTAB}. Furthermore, the TPF concept can be applied to other physical wave fields, including matter waves (e.g., an electron TPF \cite{Ehberger18PRL}). 

After more than three centuries of studying and exploiting AD, its well-established theory accounts for all the observed properties of TPFs, the most salient of which are \cite{Hall22OEConsequences}:
\begin{enumerate}
\item 
The pulse-front tilt angle for a TPF (the AD-induced angle between the phase and pulse fronts) is determined by a universal relationship that is device-independent, and is independent of the pulse bandwidth and shape \cite{Bor93OE,Hebling96OQE}.
\item
The group velocity along the propagation axis of a TPF in free space is $c$ (the speed of light in vacuum). Tuning the TPF group velocity away from $c$ requires large-angle, off-axis propagation.
\item
TPFs cannot be propagation invariant in free space because they always experience GVD along the propagation axis.
\item
A canonical result in laser physics proves that AD produces only \textit{anomalous} GVD in free space \cite{Martinez84JOSAA}. Consequently, TPFs can be exploited for GVD cancellation in materials only in their normal-GVD regime (not to be confused with normal group-delay dispersion in a Martinez compressor \cite{Martinez1987}).
\item
Individual higher-order dispersion terms experienced by TPFs in free space cannot be eliminated.
\end{enumerate}

All these effects are readily accounted for by the conventional theoretical model in which the frequency-dependent propagation angle $\varphi(\omega)$ is expanded perturbatively into a Taylor series around a fixed frequency $\omega_{\mathrm{o}}$: $\varphi(\omega)\!=\!\sum_{m}\tfrac{1}{m!}\varphi_{\mathrm{o}}^{(m)}(\omega-\omega_{\mathrm{o}})^{m}$; where $\varphi_{\mathrm{o}}^{(m)}=\tfrac{d^{m}\varphi}{d\omega^{m}}\big|_{\omega_{\mathrm{o}}}$ are the AD coefficients evaluated at a fixed frequency $\omega\!=\!\omega_{\mathrm{o}}$. As in any perturbative expansion, the impact of these AD coefficients typically drops with $m$, and most applications exploit only the first AD order $\varphi_{\mathrm{o}}^{(1)}$. The theory of space-time coupling involving only first-order effects is now fully elucidated \cite{Akturk05OE}[]. This state of affairs is justified by the fact that no known optical components provide independent and precise control over $\varphi_{\mathrm{o}}^{(2)}$ or higher-order terms. An implicit assumption underpinning the conventional theory is that the perturbative expansion for $\varphi(\omega)$ of course exists at $\omega_{\mathrm{o}}$, which requires that $\{\varphi_{\mathrm{o}}^{(m)}\}$ are all defined \cite{Porras03PRE2,Hebling06JOSAA}. In other words, the AD profile $\varphi(\omega)$ is \textit{differentiable} at $\omega_{\mathrm{o}}$ [Fig.~\ref{Fig:AngularDispersion}(b,c)].

\subsection{Why revisit the foundations of AD?}

If AD is so well-understood, why revisit the foundations of this fundamental phenomenon? Recently, the long-settled notions associated with AD have been challenged by the emergence of a new class of pulsed optical fields dubbed `space-time wave packets' (STWPs) \cite{Kondakci16OE,Parker16OE,Yessenov19OPN,Yessenov22AOP}. These are spatiotemporally structured fields undergirded by AD: each wavelength in an STWP travels at a prescribed direction -- just as is the case for TPFs. Surprisingly, the properties of STWPs \textit{violate all the rules that have been established for optical fields incorporating AD such as TPFs} \cite{Hall22OEConsequences}:
\begin{enumerate}
\item
The pulse-front tilt angle for an STWP is inversely proportional to the square-root of the bandwidth in the vicinity of a particular wavelength \cite{Hall21OL}.
\item
The group velocity of STWPs along the propagation axis can be continuously swept over an unprecedented span (below and above $c$) \cite{Kondakci19NC,Bhaduri19Optica,Bhaduri20NP} in the paraxial regime.
\item
STWPs can be entirely free of dispersion of all orders, and can thus be maintained propagation invariant in free space over extended distances \cite{Bhaduri18OE,Bhaduri19OL,Hall25OE1km}.
\item
STWPs can be designed to incur normal as well as anomalous GVD in free space \cite{Yessenov21ACSP,Hall21PRAVwave,Hall21APLP,Hall21OLNormalGVD,Hall21OLTalbot}, and thus can be used for dispersion cancellation in materials in both the anomalous- and normal-GVD regimes \cite{Hall23LPR,Hall23NP,Hall24ACSP,Hall25PRA}.
\item
Carefully sculpting the spatiotemporal spectrum of an STWP can induce arbitrary dispersion profiles never before realized in optics \cite{Yessenov21ACSP}; e.g., isolate and manipulate the coefficient (both magnitude and sign) associated with any specific dispersion order while simultaneously suppressing all others, or producing a prescribed superposition of multiple dispersion orders.
\end{enumerate}
These developments call into question the well-established rules-of-thumb regarding the consequences of AD in optics.
 
In addition, STWPs offer new modalities of interaction with photonic devices by virtue of their AD-induced field structure. For example, STWPs can be coupled to planar Fabry-P{\'e}rot cavities whose resonant linewidths are significantly narrower than the STWP bandwidth -- a phenomenon we have dubbed `omni-resonance' \cite{Shabahang17SR,Shabahang19OL,Shiri20OL,Shiri20APLP,Villinger21AOM,Shiri22OL,Layton23WhiteOmni}. Moreover, STWPs are the basis for non-dispersive `hybrid space-time modes' in planar waveguides whose propagation invariance along the unbounded waveguide dimension is maintained \cite{Shiri20NC}, and their modal indices can be tuned independently of the waveguide structure \cite{Shiri22ACSP}. In addition, propagation-invariant, group-velocity-tunable STWPs can be designed for multimode waveguides \cite{Shiri2022Optica,Shiri23JOSAA} and fibers \cite{Kibler21PRL,Guo21PRR,Bejot21ACSP,Bejot22ACSP,Stefanska23ACSP,Su25NC}. Similar characteristics are exhibited at metal-dielectric interfaces by space-time surface plasmon polaritons \cite{Schepler20ACSP}.

\subsection{Non-differentiable AD}

In light of these findings, a critical question arises: what feature of the AD inculcated into STWPs produces these departures from traditional expectations and violates the well-established limits and constraints associated with AD?

We offer here the following perspective: a new form of previously inaccessible AD underpins STWPs and produces these novel characteristics. Specifically, the AD underlying STWPs is \textit{non-differentiable}; that is, the functional dependence of the propagation angle $\varphi(\omega)$ on frequency $\omega$ is such that its derivative $\tfrac{d\varphi}{d\omega}$ is \textit{not} defined at some frequency $\omega_{\mathrm{o}}$ [Fig.~\ref{Fig:AngularDispersion}(d)], which we denote the non-differentiable frequency. Here $\varphi(\omega)$ does not correspond to a mathematically exotic structure; rather, it is finite and continuous everywhere. Moreover, $\varphi(\omega)$ is differentiable everywhere except at one frequency $\omega_{\mathrm{o}}$ where $\tfrac{d\varphi}{d\omega}|_{\omega_{\mathrm{o}}}$ is not defined. We refer to the conventional scenario as \textit{differentiable} AD, an implicit assumption always taken for granted that undergirds the perturbative expansion of $\varphi(\omega)$. In contrast, the non-differentiable AD associated with STWPs does \textit{not} possess a Taylor expansion in the vicinity of the non-differentiable frequency, whereupon the usual perturbative treatment fails.

\subsection{Universal AD synthesizer}

Common optical devices used to introduce AD typically tune only the first-order AD term $\varphi_{\mathrm{o}}^{(1)}$. Independent control over higher-order AD coefficients opens new vistas for optical physics. However, no known optical device provides such a capability, which requires constructing a new optical system: a universal AD synthesizer. This is a system that efficiently introduces an arbitrary angular profile $\varphi(\omega)$ into a generic optical field (e.g., a plane-wave pulse) with high precision. We have developed a universal AD synthesizer for shaping the spatiotemporal spectra of pulsed fields \cite{Kondakci17NP,Yessenov19OPN,Hall24JOSAA,Romer25JOpt}. Controlling the AD profile in two transverse spatial dimensions, or `conical-AD', is a more difficult task, and progress has been reported on that front only recently \cite{Yessenov22NC,Yessenov22OL,Piccardo23NP,Yessenov25Meron}. By offering ready access to arbitrary AD profiles, new problems in laser physics can be investigated and unexpected propagation behaviors can be exhibited by pulsed optical beams.

\subsection{Overview of this Perspective}

The conventional perturbative theory of differentiable AD makes a set of well-defined predictions regarding the behavior of TPFs and optical fields endowed with AD in general. After defining the central terms used throughout this article in Section~\ref{Sec:DefinitionOfTerms}, we review in Section~\ref{section:Theory of conventional angular dispersion} these predictions in nondispersive media. We also highlight a useful analogy between AD and chromatic dispersion. As a case study, we examine in Section~\ref{section:Grating} the generic example of AD induced by a diffraction grating, and comment on recent developments in metasurfaces. In Section~\ref{Section:ADControl} we explore the consequences of exercising independent control over the lowest-order AD coefficients $\varphi_{\mathrm{o}}$, $\varphi_{\mathrm{o}}^{(1)}$ and $\varphi_{\mathrm{o}}^{(2)}$. This paves the way to describing non-differentiable AD in Section~\ref{Section:NonDifferentiableAD}, where we show how it enables overturning the well-established restrictions imposed by differentiable AD. We outline in Section~\ref{Section:Classification} a classification scheme for pulsed optical fields according to their lowest AD terms, and identify which of these classes cannot be synthesized with conventional approaches. We then describe a universal AD synthesizer in Section~\ref{Section:UniversalADSynthesizer}, which is a pulsed beam shaper capable of introducing arbitrary AD profile $\varphi(\omega)$ into a plane-wave pulse. This is followed in Section~\ref{2D_NDAD} with an extension of non-differentiable AD to two transverse dimensions (conical-AD). Finally, we discuss in Section~\ref{Section:NDAD_as_a_resource} quantifying non-differentiable AD as a resource via a Schmidt number associated with the field structure, before closing with a roadmap for further developments in Section~\ref{Section:Roadmap}.


\section{Definition of terms}\label{Sec:DefinitionOfTerms}

\subsection{Angular dispersion (AD)}

Consider a scalar, plane-wave pulse $E(x,z;t)\!=\!e^{i(k_{\mathrm{o}}z-\omega_{\mathrm{o}}t)}\psi(x,z;t)$ in free space, where $x$ is the transverse coordinate, $z$ the axial coordinate, $\omega_{\mathrm{o}}$ a carrier frequency, $k_{\mathrm{o}}\!=\!\omega_{\mathrm{o}}/c$ the associated wave number. We thus do not consider here the impact of chromatic dispersion or anisotropy in the propagation medium. The angular spectrum of the slowly varying envelope $\psi(x,z;t)$ for a plane-wave pulse is independent of $x$:
\begin{equation}\label{Eq:PlaneWavePulseEnvelope}
\psi(x,z;t)=\psi(z;t)=\int\!d\Omega\widetilde{\psi}(\Omega)e^{i\{(k-k_{\mathrm{o}})z-\Omega t\}}=\psi(0;t-z/c);
\end{equation}
here $\Omega\!=\!\omega-\omega_{\mathrm{o}}$, $k\!=\!\tfrac{\omega}{c}\!=\!k_{\mathrm{o}}+\tfrac{\Omega}{c}$, and $\widetilde{\psi}(\Omega)$ is the Fourier transform of $\psi(0;t)$. All the frequencies travel in the \textit{same} direction along $z$, and the pulse propagates invariantly at a group velocity $\widetilde{v}\!=\!c$ [Fig.~\ref{Fig:AngularDispersion}(a)]. \textit{Angular dispersion} (AD) refers to field configurations in which each temporal frequency $\omega$ travels at a different angle $\varphi(\omega)$ with respect to the $z$-axis [Fig.~\ref{Fig:AngularDispersion}(b,c)]. The pulse is no longer propagation invariant, and the envelope is given by:
\begin{equation}\label{Eq:PlaneWavePulse}
\psi(x,z;t)=\int\!d\Omega\widetilde{\psi}(\Omega)e^{i\{k_{x}(\omega)x+(k_{z}(\omega)-k_{\mathrm{o}})z-\Omega t\}},
\end{equation}
where $k_{x}(\omega)\!=\!k\sin{\{\varphi(\omega)\}}$ and $k_{z}(\omega)\!=\!k\cos{\{\varphi(\omega)\}}$ are the transverse and axial wave numbers, respectively, and $\widetilde{\psi}(\Omega)$ is the Fourier transform of $\psi(0,0;t)$. We return to the question of introducing conical-AD in Section~\ref{2D_NDAD}.

\subsection{Group-velocity dispersion (GVD) versus group-delay dispersion (GDD)}

\begin{figure}[t!]
\centering
\includegraphics[width=8.6cm]{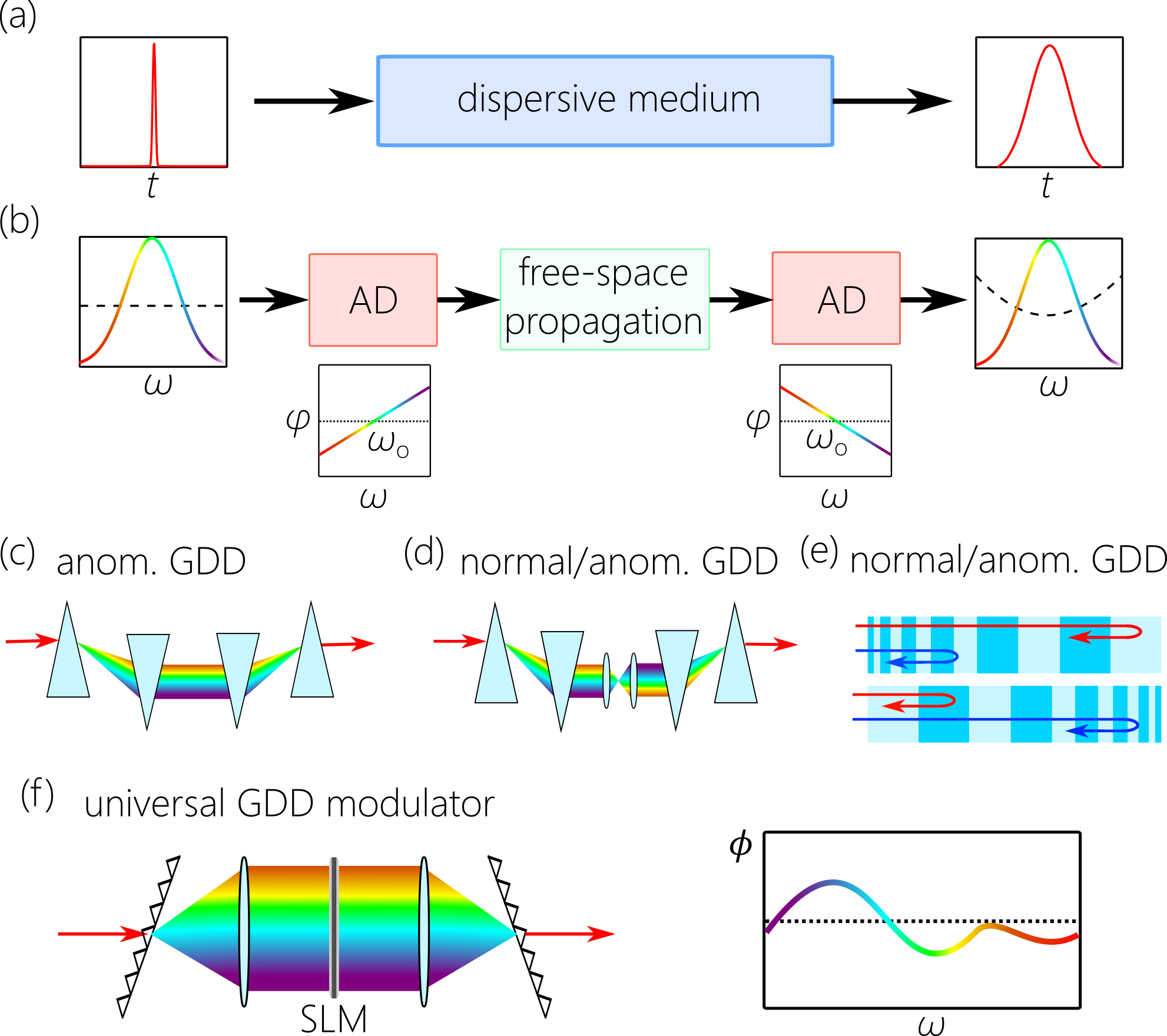}
\caption{GDD is introduced into a transform-limited pulse (a) after traversing a dispersive medium, or (b) by introducing AD, free propagation, then removing the AD. The continuous and dashed curves are the spectral amplitudes and phase, respectively. (c-f) Examples of systems that introduce GDD: (c) a sequence of prisms (anom. = anomalous); (d) a Martinez compressor; or (e) chirped Bragg gratings; and (f) a universal GDD modulator based on spectral-phase modulation with a spatial light modulator (SLM). Here $\phi$ is the spectral phase.}
\label{Fig:GDD}
\end{figure}

It is crucial to distinguish here between GVD and GDD. A pulse gradually broadens temporally upon traversing a dispersive medium (chromatic dispersion) as a result of GVD [Fig.~\ref{Fig:GDD}(a)], which corresponds to an additional quadratic phase term $e^{ik_{2}\Omega^{2}z/2}$ introduced into the angular spectrum in Eq.~\ref{Eq:PlaneWavePulseEnvelope}; where $k_{2}$ is the GVD coefficient that can be positive (normal GVD) or negative (anomalous GVD) \cite{SalehBook07}. The \textit{final} pulse broadening after the dispersive medium is quantified by the GDD, which is the second-order derivative of the accumulated spectral phase $\phi(\omega)$, $\tfrac{d^{2}\phi}{d\omega^{2}}\big|_{\omega_{\mathrm{o}}}\!\sim\!k_{2}L$, where $L$ is the medium length.

\begin{figure}[t!]
\centering
\includegraphics[width=8.6cm]{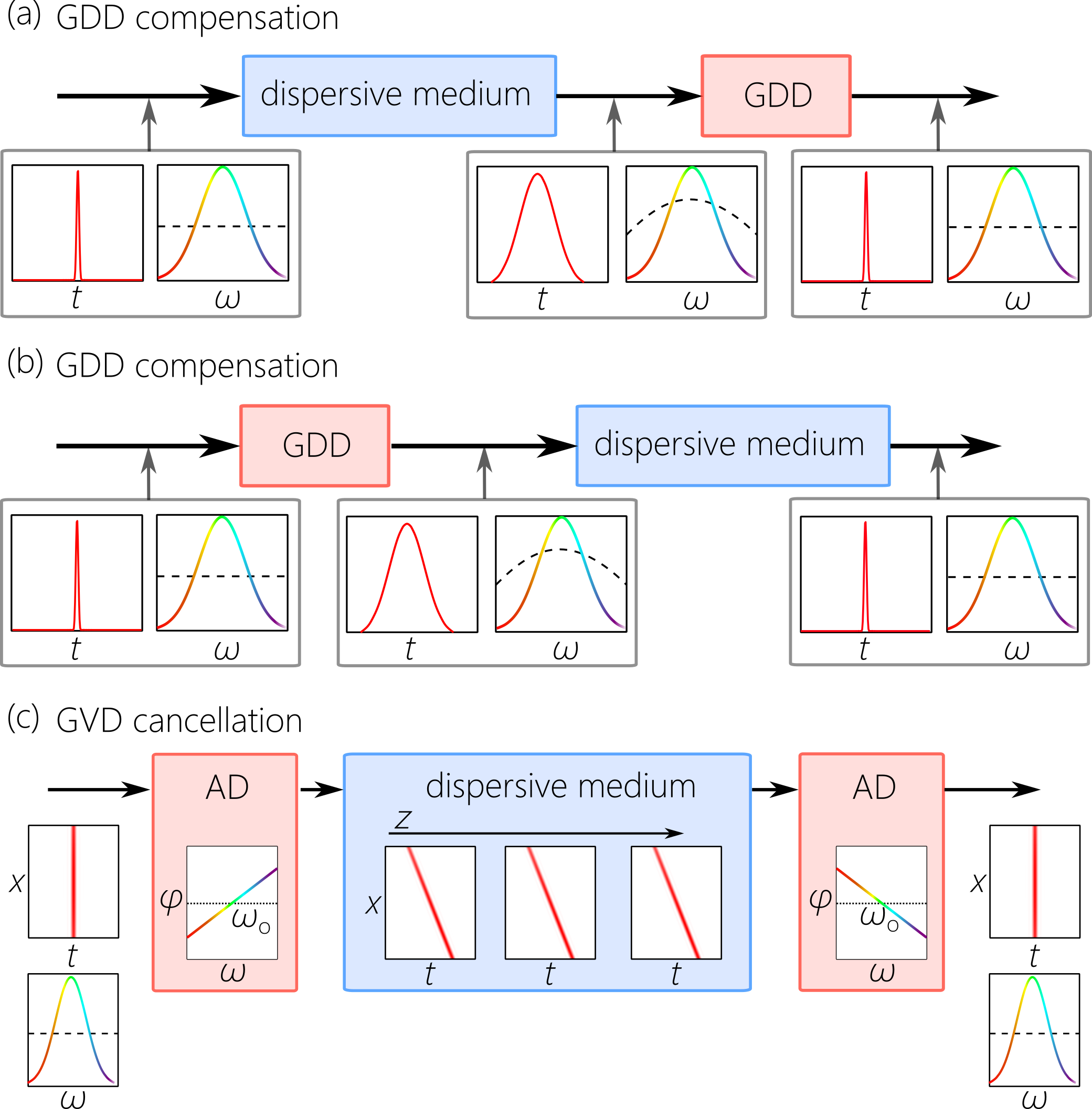}
\caption{(a,b) GDD compensation. (a) After traversing a dispersive medium, the GDD accumulated can be compensated for by introducing GDD [Fig.~\ref{Fig:GDD}] of equal magnitude but opposite sign. (b) Conversely, GDD can be pre-compensated for. (c) GVD cancellation. A field endowed with AD experiences GVD in free space, and can thus propagate across a dispersive medium invariantly. Both the spatiotemporal profile and complex spectrum at the output thus coincide with those at the input.}
\label{Fig:GDDCompensation}
\end{figure}

 \begin{figure*}[t!]
\centering
\includegraphics[width=17.6cm]{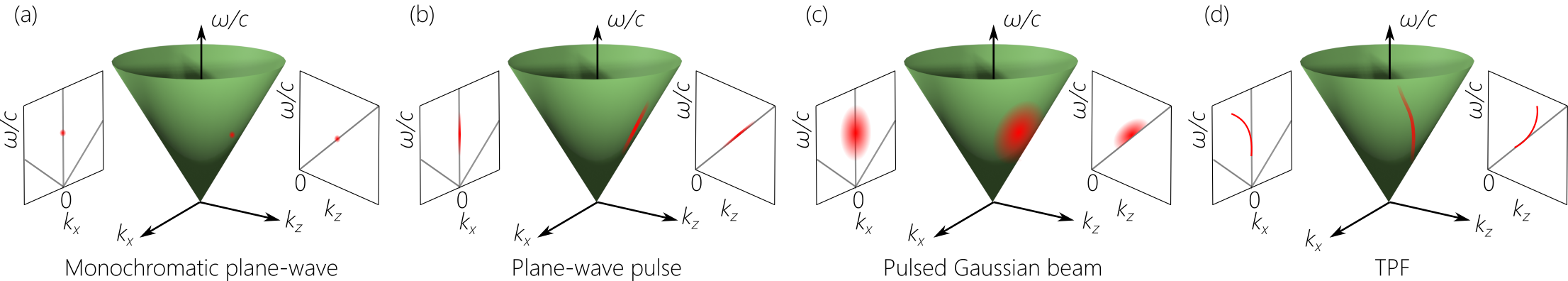}
\caption{The spectral support for optical fields on the surface of the free-space light-cone $k_{x}^{2}+k_{z}^{2}\!=\!(\tfrac{\omega}{c})^{2}$. The dotted lines are the light-lines. (a) A monochromatic plane wave is represented by a point; (b) a plane-wave pulse traveling along the $z$-axis by the straight-line $k_{z}\!=\!\omega/c$; (c) a conventional pulsed beam by a 2D domain; and (d) a tilted pulse front (TPF) by a 1D curved trajectory.}
\label{Fig:LightCone}
\end{figure*}

Introducing AD into a plane-wave pulse induces GVD in free space [Fig.~\ref{Fig:GDD}(b)], just as in a dispersive medium. After removing the AD, the retrieved plane-wave pulse no longer experiences GVD, and the final GDD value is subsequently retained. The distinction between GVD and GDD should now be clear: GVD is a distributed effect corresponding to the gradual accumulation of spectral phase and pulse broadening via either chromatic dispersion in a medium or AD in free space; whereas GDD is the final spectral-phase accumulated after undergoing GVD. Multiple approaches to realize AD-induced GDD are illustrated in Fig.~\ref{Fig:GDD}(c-f), such as a sequence of prisms [Fig.~\ref{Fig:GDD}(c)]; a Martinez compressor [Fig.~\ref{Fig:GDD}(d)]; and chirped Bragg gratings [Fig.~\ref{Fig:GDD}(e)]. A well-known theorem put forth by Martinez, Gordon, and Fork (henceforth MGF) \cite{Martinez84JOSAA} shows that a pulse after introducing AD can experience \textit{only} anomalous GVD in free space, and thus accumulate only anomalous GDD. We shall return to this theorem and reassess its claim in Section~\ref{section:Theory of conventional angular dispersion}. Nevertheless, normal or anomalous GDD can be produced in a Martinez compressor \cite{Martinez1987} via AD [Fig.~\ref{Fig:GDD}(d)]. Even though the TPF in each segment of the compressor experiences anomalous GVD, the overall system can introduce anomalous or normal GDD because of the spatially inverting lens system. A variety of specialized systems have been developed for producing higher-order GDD \cite{Lemoff93OL,White93OL,Kane97JOSAB,Kane97JOSAB2}. Finally, the standard pulse-shaper in Fig.~\ref{Fig:GDD}(f) is a \textit{universal GDD modulator}, because an arbitrary spectral phase $\phi(\omega)$ can be added by a spatial-light modulator (SLM) to produce either normal or anomalous GDD by changing the sign of the spectral phase, or produce higher-order dispersion terms \cite{Weiner00RSI}.

\subsection{GDD compensation versus GVD cancellation}

In coherent pulse amplification, an ultrashort pulse is chirped and amplified, before introducing oppositely signed GDD to return the pulse to its initial width. We refer to this procedure as \textit{GDD compensation} [Fig.~\ref{Fig:GDDCompensation}(a,b)]. We distinguish between GDD compensation and another procedure we call \textit{GVD cancellation} [Fig.~\ref{Fig:GDDCompensation}(c)], whereby a pulsed field traverses a dispersive medium \textit{without} temporal broadening after initially introducing AD. Such a pulsed field would encounters GVD in free space [Fig.~\ref{Fig:GDD}(b)]. However, once coupled to a medium of equal GVD magnitude but opposite sign, it propagates invariantly and emerges with its initial width intact. According to the MGF theorem \cite{Martinez84JOSAA}, AD can introduce only anomalous GVD in free space, and thus yields GVD cancellation only in the normal-GVD regime.

\subsection{Spectral representation on the light-cone}

A useful tool for visualizing field configurations incorporating AD is their spectral representation on the surface of the light-cone $k_{x}^{2}+k_{z}^{2}\!=\!(\tfrac{\omega}{c})^{2}$ in Fig.~\ref{Fig:LightCone} \cite{Donnelly93ProcRSLA,Saari04PRE}. Because most devices that introduce AD (e.g., gratings and prisms) do so in one transverse dimension only, we ignore the second transverse dimension $y$ (we return to this question and generalize AD to two transverse dimensions in Section~\ref{2D_NDAD}). A monochromatic plane wave $e^{i(k_{x}x+k_{z}z-\omega t)}$ is represented by a \textit{point} on the light-cone surface [Fig.~\ref{Fig:LightCone}(a)], and any propagating field thus corresponds to some domain on the light-cone. For example, the plane-wave pulse in Eq.~\ref{Eq:PlaneWavePulseEnvelope} [Fig.~\ref{Fig:AngularDispersion}(a)] is represented on the light-line $k_{z}\!=\!\tfrac{\omega}{c}$ associated with $k_{x}\!=\!0$ [Fig.~\ref{Fig:LightCone}(b)], whereas the spectral support for a conventional pulsed beam of finite spatial and temporal bandwidths is a two-dimensional (2D) domain [Fig.~\ref{Fig:LightCone}(c)]. The spectral support for a TPF incorporating AD [Eq.~\ref{Eq:PlaneWavePulse} and Fig.~\ref{Fig:AngularDispersion}(b,c)] is a one-dimensional (1D) curved trajectory because $k_{x}$ and $k_{z}$ are both dependent on $\omega$ [Fig.~\ref{Fig:LightCone}(d)]. Although a TPF has finite spatial and temporal bandwidths just like a conventional pulsed beam, the spectral support is nevertheless a 1D trajectory rather than a 2D domain, because of the association between the temporal frequencies $\omega$ and the wave-vector components $k_{x}(\omega)$ and $k_{z}(\omega)$. We are concerned hereon with fields incorporating AD whose spectral supports are 1D curves on the light-cone surface. 

\section{Conventional angular dispersion}
\label{section:Theory of conventional angular dispersion}

The conventional theory of AD expands the propagation angle $\varphi(\omega)$ around a fixed frequency $\omega\!=\!\omega_{\mathrm{o}}$:
\begin{equation}\label{Eq:TaylorSeriesForPropagationAngle}
\varphi(\omega)\!=\!\varphi(\omega_{\mathrm{o}}+\Omega)\!\approx\!\varphi_{\mathrm{o}}+\varphi_{\mathrm{o}}^{(1)}\Omega+\frac{1}{2}\varphi_{\mathrm{o}}^{(2)}\Omega^{2}+\cdots,
\end{equation}
where $\varphi_{\mathrm{o}}\!=\!\varphi(\omega_{\mathrm{o}})$, $\varphi_{\mathrm{o}}^{(n)}\!=\!\tfrac{d^{n}\varphi}{d\omega^{n}}\big|_{\omega=\omega_{\mathrm{o}}}$, and we refer to $\{\varphi_{\mathrm{o}}^{(n)}\}$ as the AD coefficients. The subscript `o' indicates quantities evaluated at $\omega_{\mathrm{o}}$. The transverse and axial wave numbers in Eq.~\ref{Eq:PlaneWavePulse} are in turn expanded as follows:
\begin{equation}\label{Eq:kxkz}
k_{x}(\omega_{\mathrm{o}}+\Omega)\!=\!\sum_{n=0}^{\infty}\frac{1}{n!}k_{x}^{(n)}\Omega^{n},\;\;\; k_{z}(\omega_{\mathrm{o}}+\Omega)\!=\!\sum_{n=0}^{\infty}\frac{1}{n!}k_{z}^{(n)}\Omega^{n}.
\end{equation}
We refer to the set of coefficients $\{k_{x}^{(n)}\}$ and $\{k_{z}^{(n)}\}$ in these two expansions as the transverse and axial dispersion coefficients, respectively; $\{k_{x}^{(n)}\}$ account for transverse effects such as transverse walk-off, while $\{k_{z}^{(n)}\}$ determine the axial propagation dynamics. We adopt throughout a normalization scheme with dimensionless coefficients $\omega_{\mathrm{o}}^{n}\varphi_{\mathrm{o}}^{(n)}$, $c\omega_{\mathrm{o}}^{n-1}k_{x}^{(n)}$, and $c\omega_{\mathrm{o}}^{n-1}k_{z}^{(n)}$. Equations~\ref{Eq:PlaneWavePulse}, \ref{Eq:TaylorSeriesForPropagationAngle}, and \ref{Eq:kxkz} form the basis for the conventional theory of AD.

\begin{figure*}[t!]
\centering
\includegraphics[width=17.6cm]{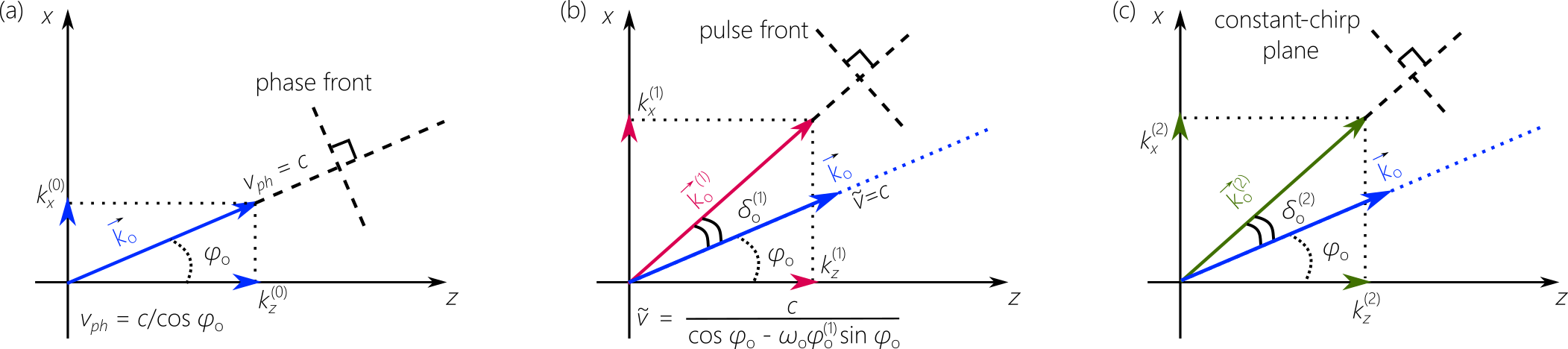}
\caption{(a) The phase front is orthogonal to $\vec{k}_{\mathrm{o}}$, $v_{\mathrm{ph}}\!=\!c$ along $\vec{k}_{\mathrm{o}}$, whereas $v_{\mathrm{ph}}\!=\!c/\cos{\varphi_{\mathrm{o}}}$ along the $z$-axis. (b) The pulse front is orthogonal to $\vec{k}_{\mathrm{o}}^{(1)}$, which makes an angle $\delta_{\mathrm{o}}^{(1)}$ with $\vec{k}_{\mathrm{o}}$. Along $\vec{k}_{\mathrm{o}}$, $\widetilde{v}=c$; along the $z$-axis, $\widetilde{v}$ is given by Eq.~\ref{Eq:GroupVelocityAngularDispersion}; and along $\vec{k}_{\mathrm{o}}^{(1)}$, $\widetilde{v}$ is given by Eq.~\ref{eq:GroupVelAlongk1}. (c) The plane of constant chirp is orthogonal to the vector $\vec{k}_{\mathrm{o}}^{(2)}$ (Eq.~\ref{Eq:SecondOrderDispersionCoefficients}), which makes an angle $\delta_{\mathrm{o}}^{(2)}$ with $\vec{k}_{\mathrm{o}}$.}
\label{Fig:DeltaAngles}
\end{figure*}

In this formulation, $\omega_{\mathrm{o}}$ travels at an angle $\varphi_{\mathrm{o}}$ with the $z$-axis, where $\varphi_{\mathrm{o}}$ is the central angle of the angular span of the field [Fig.~\ref{Fig:DeltaAngles}(a)]. We refer to the case where $\varphi_{\mathrm{o}}=0$ (i.e., $\omega_{\mathrm{o}}$ travels along $z$) as an `on-axis' field, and $\varphi_{\mathrm{o}}\neq0$ as an `off-axis' field. The equations for the dispersion coefficients are simplified for on-axis fields, and we denote the dispersion coefficients in this case $\{k_{x}^{(n)}(0)\}$ and $\{k_{z}^{(n)}(0)\}$. However, in many scenarios it is more useful or convenient to adopt a different `observation axis', a direction along which a particular nonlinear interaction is enabled or one that is orthogonal to a surface of interest. Rotating the coordinate system in the $(x,z)$-plane by an angle $\varphi_{\mathrm{o}}$ yields:
\begin{eqnarray}\label{Eq:GeometricRotation}
c\omega_{\mathrm{o}}^{n-1}k_{x}^{(n)}(\varphi_{\mathrm{o}})\!\!\!&=&\!\!\!c\omega_{\mathrm{o}}^{n-1}k_{x}^{(n)}(0)\cos{\varphi_{\mathrm{o}}}+c\omega_{\mathrm{o}}^{n-1}k_{z}^{(n)}(0)\sin{\varphi_{\mathrm{o}}},\nonumber\\
c\omega_{\mathrm{o}}^{n-1}k_{z}^{(n)}(\varphi_{\mathrm{o}})\!\!\!&=&\!\!\!c\omega_{\mathrm{o}}^{n-1}k_{z}^{(n)}(0)\cos{\varphi_{\mathrm{o}}}-c\omega_{\mathrm{o}}^{n-1}k_{x}^{(n)}(0)\sin{\varphi_{\mathrm{o}}}.
\end{eqnarray}
Whereas a coordinate rotation does \textit{not} affect the values of the AD coefficients $\{\varphi_{\mathrm{o}}^{(n)}\}$, of course with the exception of $\varphi_{\mathrm{o}}$, it \textit{does} change the values of the dispersion coefficients $\{k_{x}^{(n)}\}$ and $\{k_{z}^{(n)}\}$. Such transformations are a versatile tool for producing a desired dispersion profile.

\subsection{Phase front and phase velocity}

The phase front (the plane of constant \textit{phase}) is orthogonal to $\vec{k}_{\mathrm{o}}\!=\!k_{x}^{(0)}\hat{x}+k_{z}^{(0)}\hat{z}$, where $k_{x}^{(0)}\!=\!k_{\mathrm{o}}\sin{\varphi_{\mathrm{o}}}$ and $k_{z}^{(0)}\!=\!k_{\mathrm{o}}\cos{\varphi_{\mathrm{o}}}$  [Fig.~\ref{Fig:DeltaAngles}(a)]. The phase velocity is $v_{\mathrm{ph}}\!=\!\omega_{\mathrm{o}}/k_{\mathrm{o}}\!=\!c$ along $\vec{k}_{\mathrm{o}}$ in free space, but along the $z$-axis the phase velocity is $v_{\mathrm{ph}}\!=\!c/\cos{\varphi_{\mathrm{o}}}\!>\!c$. This deviation of the phase velocity from $c$ is a \textit{geometric} effect, similarly to the phenomenon arising commonly in the context of monochromatic cosine waves $\cos{(k_{\mathrm{o}}x\sin{\varphi_{\mathrm{o}}})}$ or Bessel beams $J_{m}(k_{\mathrm{o}}r\sin{\varphi_{\mathrm{o}}})$ for integer $m$. Both beams comprise plane waves all making the \textit{same} angle $\varphi_{\mathrm{o}}$ with the $z$-axis, resulting in an axial phase velocity $c/\cos{\varphi_{\mathrm{o}}}$ rather than $c$ \cite{Chiao02OPN,Milonni02JPB}.

\subsection{Pulse front and group velocity}

By aligning $\vec{k}_{\mathrm{o}}$ with the $z$-axis ($\varphi_{\mathrm{o}}\!=\!0$), the first-order dispersion coefficients are $ck_{x}^{(1)}(0)\!=\!\omega_{\mathrm{o}}\varphi_{\mathrm{o}}^{(1)}$ and $ck_{z}^{(1)}(0)\!=\!1$. The pulse front (the plane of constant \textit{intensity}) is normal to $\vec{k}_{\mathrm{o}}^{(1)}\!=\!k_{x}^{(1)}x+k_{z}^{(1)}z$, which makes an angle $\delta_{\mathrm{o}}^{(1)}$ with $\vec{k}_{\mathrm{o}}$ [Fig.~\ref{Fig:DeltaAngles}(b)], where:
\begin{equation}\label{Eq:AngleOfPulseFrontTilt}
\tan{\delta}_{\mathrm{o}}^{(1)}=\omega_{\mathrm{o}}\varphi_{\mathrm{o}}^{(1)}.
\end{equation}
This well-established relationship for TPFs is considered `universal': the pulse-front tilt angle $\delta_{\mathrm{o}}^{(1)}$ is device-independent, is independent of the pulse shape and of its bandwidth, and depends solely on first-order AD \cite{Bor93OE,Hebling96OQE}. The field profile can be symmetrized when $k_{x}(\omega)$ assumes symmetric positive and negative values, leading to the overlap of two TPFs rotated by $\pm\delta_{\mathrm{o}}^{(1)}$ with respect to $\vec{k}_{\mathrm{o}}$ \cite{Kondakci19ACSP}, in which case its spatiotemporal profile is X-shaped, thus resembling that of X-waves \cite{Porras03PRE2} and STWPs \cite{Kondakci19ACSP}. The angle between $\vec{k}_{\mathrm{o}}$ and $\vec{k}_{\mathrm{o}}^{(1)}$ is not a result of anisotropy in the medium, but stems solely from AD.

Whereas $v_{\mathrm{ph}}$ is affected by geometric factors alone, the group velocity $\widetilde{v}$ receives both geometric \textit{and} interferometric contributions; that is, $\widetilde{v}$ depends on both $\varphi_{\mathrm{o}}$ and $\varphi_{\mathrm{o}}^{(1)}$. We assume throughout that the $z$-axis is the observation axis. When $\vec{k}_{\mathrm{o}}$ lies along the $z$-axis ($\varphi_{\mathrm{o}}=0$), then $\widetilde{v}=c=v_{\mathrm{ph}}$; when $\vec{k}_{\mathrm{o}}^{(1)}$ coincides with the $z$ axis ($\varphi_{\mathrm{o}}\!=\!-\delta_{\mathrm{o}}^{(1)}$),
\begin{equation}\label{eq:GroupVelAlongk1}
\widetilde{v}=c\cos{\varphi_{\mathrm{o}}}=\frac{c}{\sqrt{1+(\omega_{\mathrm{o}}\varphi_{\mathrm{o}}^{(1)})^{2}}},
\end{equation}
which is always subluminal, along with a superluminal phase velocity $v_{\mathrm{ph}}\!=\!\tfrac{c}{\cos{\varphi_{\mathrm{o}}}}$, so that $v_{\mathrm{ph}}\widetilde{v}=c^{2}$; when $\vec{k}_{\mathrm{o}}$ makes an angle $\varphi_{\mathrm{o}}$ with the $z$-axis, then $\widetilde{v}$ along $z$ is:
\begin{equation}\label{Eq:GroupVelocityAngularDispersion}
\widetilde{v}=\frac{c}{\cos{\varphi_{\mathrm{o}}}-\omega_{\mathrm{o}}\varphi_{\mathrm{o}}^{(1)}\sin{\varphi_{\mathrm{o}}}}=c\frac{\cos{\delta_{\mathrm{o}}^{(1)}}}{\cos{(\delta_{\mathrm{o}}^{(1)}\!+\!\varphi_{\mathrm{o}})}}=\frac{c}{\widetilde{n}},
\end{equation}
where $\widetilde{n}$ is the group index. 

In principle, $\widetilde{v}$ can take on arbitrary values by judicious choice of $\varphi_{\mathrm{o}}^{(1)}$ (an intrinsic interferometric factor) and $\varphi_{\mathrm{o}}$ (an extrinsic geometric factor), including even negative values of $\widetilde{v}$. Negative $\widetilde{n}\!<\!0$ is realized when $\tan{\delta_{\mathrm{o}}^{(1)}}\tan{\varphi_{\mathrm{o}}}\!>\!1$ or $\cos(\delta_{\mathrm{o}}^{(1)}\!+\!\varphi_{\mathrm{o}})\!<\!0$, which entails that $\delta_{\mathrm{o}}^{(1)}\!+\!\varphi_{\mathrm{o}}\!>\!90^{\circ}$ \cite{Zapata06OL}. Broad tunability of $\widetilde{v}$ at small angles $\varphi_{\mathrm{o}}$ requires large values of $\omega_{\mathrm{o}}\varphi_{\mathrm{o}}^{(1)}$. For example, if $\omega_{\mathrm{o}}\varphi_{\mathrm{o}}^{(1)}\!=\!50$ ($\delta_{\mathrm{o}}^{(1)}\!\approx\!88.85^{\circ}$), then tuning $\varphi_{\mathrm{o}}$ from $0^{\circ}$ to $3^{\circ}$ provides a large swing in the value of $\widetilde{v}$ along the $z$-axis: $\widetilde{v}\!=\!c$ at $\varphi_{\mathrm{o}}\!=\!0^{\circ}$; $\widetilde{v}\!=\!7.86c$ at $\varphi_{\mathrm{o}}\!=\!1^{\circ}$; $\widetilde{v}\!\rightarrow\!\infty$ at $\varphi_{\mathrm{o}}\!\approx\!1.15^{\circ}$; after which $\widetilde{v}$ becomes negative, reaching $\widetilde{v}\!=\!-1.3c$ at $\varphi_{\mathrm{o}}\!=\!2^{\circ}$. Such a value for $\omega_{\mathrm{o}}\varphi_{\mathrm{o}}^{(1)}$ is significantly larger than that produced with common optical devices (see Section~X) Alternatively, at realistic values of AD $\omega_{\mathrm{o}}\varphi_{\mathrm{o}}^{(1)}\sim1$, a very large $\varphi_{\mathrm{o}}$ is needed (off-axis propagation), and thus control over $\widetilde{v}$ is realized along over a short interaction distance.

\subsection{Comparison of AD and chromatic dispersion}

It is instructive to compare $\widetilde{v}$ in Eq.~\ref{Eq:GroupVelocityAngularDispersion} to the group velocity $v_{\mathrm{g}}$ in a material with chromatic dispersion $n\!=\!n(\omega)$ \cite{SalehBook07},
\begin{equation}\label{Eq:GroupVelocityChromaticDispersion}
v_{\mathrm{g}}=\frac{c}{n_{\mathrm{o}}+\omega_{\mathrm{o}}n_{\mathrm{o}}^{(1)}}=\frac{c}{n_{\mathrm{g}}},
\end{equation}
where $n_{\mathrm{o}}\!=\!n(\omega_{\mathrm{o}})$, $n_{\mathrm{o}}^{(1)}\!=\!\tfrac{dn}{d\omega}\big|_{\omega_{\mathrm{o}}}$, and $n_{\mathrm{g}}$ is the group index; compare to Eq.~\ref{Eq:GroupVelocityAngularDispersion}. In contrast to the scenario of AD, deviation of $v_{\mathrm{g}}$ from $c$ is a material property. In absence of chromatic dispersion, the group velocity tends to the phase velocity $v_{\mathrm{g}}\!\rightarrow\!c/n_{\mathrm{o}}$ and $n_{\mathrm{g}}\rightarrow n_{\mathrm{o}}$. Analogously, in absence of AD in Eq.~\ref{Eq:GroupVelocityAngularDispersion}, the group velocity also trends towards the phase velocity $\widetilde{v}\!\rightarrow\!c/\cos{\varphi_{\mathrm{o}}}$. \textit{Chromatic} dispersion shifts the group index to $n_{\mathrm{o}}+\omega_{\mathrm{o}}n_{\mathrm{o}}^{(1)}$ via an interference effect in which each frequency $\omega$ acquires a different spectral phase \cite{Guo06PRE}. Equation~\ref{Eq:GroupVelocityChromaticDispersion} is the basis for `slow-light' and `fast-light' \cite{Hau99Nature,Chiao02OPN,Milonni02JPB,Baba08NP,Boyd09Science,Tsakmakidis17Science} whereby $v_{\mathrm{g}}$ can deviate substantially from $c/n_{\mathrm{o}}$ because of large chromatic dispersion $|\omega_{\mathrm{o}}n_{\mathrm{o}}^{(1)}|$. Analogously, AD shifts the group index $\cos{\varphi_{\mathrm{o}}}\!\rightarrow\!\cos{\varphi_{\mathrm{o}}}-\omega_{\mathrm{o}}\varphi_{\mathrm{o}}^{(1)}\sin{\varphi_{\mathrm{o}}}$ via a combination of interference and geometric effects. A basic distinction remains between the cases of chromatic dispersion and AD: whereas $n_{\mathrm{o}}^{(1)}$ is constrained by the properties of existing optical materials or photonic structures, $\varphi_{\mathrm{o}}^{(1)}$ can -- in principle -- be controlled arbitrarily.

\subsection{GVD and higher-order dispersion terms}

The second-order dispersion coefficients when $\varphi_{\mathrm{o}}\!=\!0$ are:
\begin{equation}\label{Eq:SecondOrderDispersionCoefficients}
c\omega_{\mathrm{o}}k_{x}^{(2)}(0)\!=\!\omega_{\mathrm{o}}^{2}\varphi_{\mathrm{o}}^{(2)}\!+\!2\omega_{\mathrm{o}}\varphi_{\mathrm{o}}^{(1)},\;\;\;
c\omega_{\mathrm{o}}k_{z}^{(2)}(0)\!=\!-(\omega_{\mathrm{o}}\varphi_{\mathrm{o}}^{(1)})^{2},
\end{equation}
the latter of which determines the axial AD-induced GVD experienced by the wave packet in free space. The plane orthogonal to the vector $\vec{k}_{\mathrm{o}}^{(2)}\!=\!k_{x}^{(2)}x+k_{z}^{(2)}z$ is a surface of constant spectral chirp, which makes an angle $\delta_{\mathrm{o}}^{(2)}$ with $\vec{k}_{\mathrm{o}}$ [Fig.~\ref{Fig:DeltaAngles}(c)]:
\begin{equation}
\tan{\delta_{\mathrm{o}}^{(2)}}=-\frac{\omega_{\mathrm{o}}^{2}\varphi_{\mathrm{o}}^{(2)}+2\omega_{\mathrm{o}}\varphi_{\mathrm{o}}^{(1)}}{(\omega_{\mathrm{o}}\varphi_{\mathrm{o}}^{(1)})^{2}}.
\end{equation}
Maximum GVD is encountered by the wave packet along this direction \cite{Porras03PRE2}. Higher-order dispersion terms can be readily derived; e.g., the third-order coefficients when $\varphi_{\mathrm{o}}\!=\!0$ are:
\begin{eqnarray}\label{Eq:ThirdOrderDispersionCoefficientsGeneral}
c\omega_{\mathrm{o}}^{2}k_{x}^{(3)}(0)\!\!\!\!&=&\!\!\!\!\omega_{\mathrm{o}}^{3}\varphi_{\mathrm{o}}^{(3)}+3\omega_{\mathrm{o}}^{2}\varphi_{\mathrm{o}}^{(2)},\nonumber\\
c\omega_{\mathrm{o}}^{2}k_{z}^{(3)}(0)\!\!\!\!&=&\!\!\!\!-3\omega_{\mathrm{o}}\varphi_{\mathrm{o}}^{(1)}(\omega_{\mathrm{o}}^{2}\varphi_{\mathrm{o}}^{(2)}+\omega_{\mathrm{o}}\varphi_{\mathrm{o}}^{(1)}).
\end{eqnarray}
Expressions for even higher-order terms can be readily obtained, but become progressively more complex.

\subsection{The GVD theorem of Martinez, Gordon, and Fork}
\label{subsection:MGF theorem}

The MGF theorem \cite{Martinez84JOSAA} makes use of Eq.~\ref{Eq:SecondOrderDispersionCoefficients}: the GVD encountered in free space after inculcating AD is always \textit{anomalous} along $\vec{k}_{\mathrm{o}}$, $c\omega_{\mathrm{o}}k_{z}^{(2)}(0)\!=\!-\tan^{2}{\delta_{\mathrm{o}}^{(1)}}$, and it can thus be used for GVD-cancellation in the normal-GVD regime of a dispersive medium. This result is \textit{not} as general as commonly thought, but is rather based on two assumptions: (1) on-axis propagation $\varphi_{\mathrm{o}}\!=\!0$, and (2) the differentiability of the AD profile, so that the perturbative expansion for $\varphi(\omega)$ is valid at $\omega_{\mathrm{o}}$, and $\varphi_{\mathrm{o}}^{(1)}$ is well-defined. Overturning either of these assumptions may lead to a violation of the conclusion, in which case AD can induce \textit{normal} GVD in free space, and thus enable GVD-cancellation in the medium's anomalous-GVD regime (see Section~\ref{Section:ADControl}). Two pathways to violating the MGF theorem are thus available. First, an \textit{off-axis} field can experience either normal or anomalous GVD along $z$, but this requires exercising independent control over $\varphi_{\mathrm{o}}$, $\varphi_{\mathrm{o}}^{(1)}$, and $\varphi_{\mathrm{o}}^{(2)}$ (Section~\ref{Section:ADControl}). No single optical component offers this capability to date. Second, non-differentiable AD can yield either normal or anomalous GVD in free space. Both of these pathways require a universal AD synthesizer. 


\begin{figure}[t!]
\centering
\includegraphics[width=8.6cm]{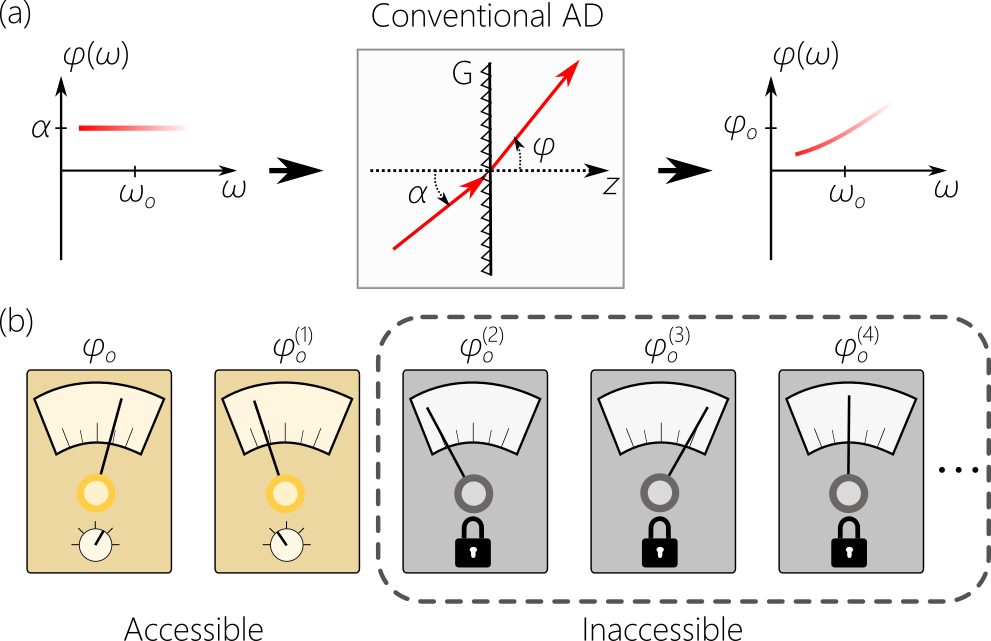}
\caption{(a) Conventional AD is introduced into a collimated pulse via a grating G, where $\alpha$ is the incident angle with respect to its normal, and $\varphi$ is the diffraction angle. The arrows for $\alpha$ and $\varphi$ indicate the direction for positive angles. (b) Only a subset of the AD coefficients are independently accessible, shown as `active' yellow knobs for $\varphi_{\mathrm{o}}^{(0)}$ and $\varphi_{\mathrm{o}}^{(1)}$, whereas all the other higher-order coefficients are not accessible independently (depicted as `locked' gray knobs).}
\label{Fig:DiffAD}
\end{figure}

\section{Case study of angular dispersion}\label{section:Grating}

\subsection{AD produced by a diffraction grating}

\begin{figure*}[t!]
\centering
\includegraphics[width=17.6cm]{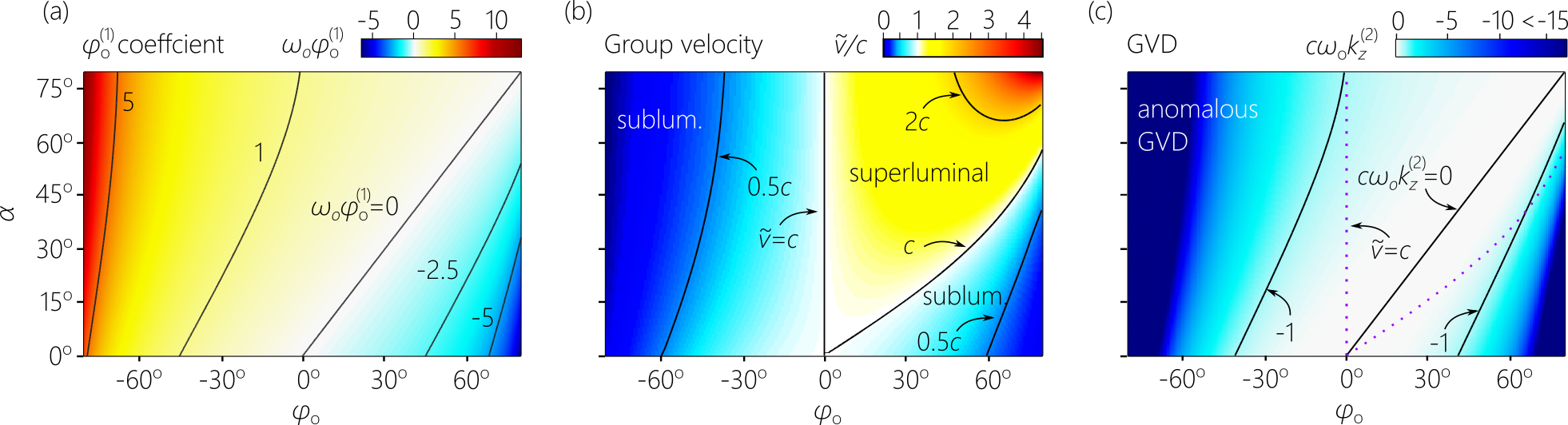}
\caption{(a) The first-order AD coefficient $\omega_{\mathrm{o}}\varphi_{\mathrm{o}}^{(1)}$ of a conventional diffraction grating as a function of input angle $\alpha$ and diffraction angle $\varphi_{\mathrm{o}}$ [Fig.~\ref{Fig:DiffAD}]. (b) The axial group velocity $\widetilde{v}$ of the TPF produced by a grating. The TPF is superluminal between the $\widetilde{v}\!=\!c$ contours, and is subluminal elsewhere. (c) The axial GVD coefficient $c\omega_{\mathrm{o}}k_{z}^{(2)}$ associated with the TPFs in (b), which is anomalous everywhere. The dotted contours correspond to luminal TPFs $\widetilde{v}\!=\!c$. The straight line along which $k_{z}^{(2)}\!=\!0$ corresponds to the absence of a grating, $\alpha\!=\!\varphi_{\mathrm{o}}$.}
\label{Fig:GratingCharacteristics}
\end{figure*}

We elucidate these general precepts regarding AD by applying them to a concrete example: a diffraction grating \cite{Rittenhouse1786TAPS,Fraunhofer1823GAN,Harvey19OE}. If the incident and diffracted angles with respect to the grating normal are $\alpha$ and $\varphi$ [Fig.~\ref{Fig:DiffAD}(a)], respectively, then $\sin{\alpha}+m\tfrac{\lambda}{\Lambda}\!=\!\sin{\varphi}$, where $\Lambda$ is the grating ruling period, $m$ is the diffraction order, and $m\tfrac{\lambda_{\mathrm{o}}}{\Lambda}\!=\!\sin{\varphi_{\mathrm{o}}}-\sin{\alpha}$ at $\omega\!=\!\omega_{\mathrm{o}}$. The grating is associated with only two independent parameters: $\alpha$ and $m\tfrac{\lambda_{\mathrm{o}}}{\Lambda}$, or equivalently $\alpha$ and $\varphi_{\mathrm{o}}$. We thus obtain:
\begin{equation}\label{Eq:ND_grating}
\omega_{\mathrm{o}}\varphi_{\mathrm{o}}^{(1)}=\frac{\sin{\alpha}-\sin{\varphi_{\mathrm{o}}}}{\cos{\varphi_{\mathrm{o}}}},\;\;
\omega_{\mathrm{o}}^{2}\varphi_{\mathrm{o}}^{(2)}=\omega_{\mathrm{o}}\varphi_{\mathrm{o}}^{(1)}\{\omega_{\mathrm{o}}\varphi_{\mathrm{o}}^{(1)}\tan{\varphi_{\mathrm{o}}}-2\}.
\end{equation}
The higher-order AD coefficients $\{\varphi_{\mathrm{o}}^{(n)}\}$ with $n\!\geq\!1$ are \textit{all} determined by $\alpha$ and $\varphi_{\mathrm{o}}$. We depict this state of affairs pictorially in Fig.~\ref{Fig:DiffAD}(b), where the two lowest-order AD coefficients $\varphi_{\mathrm{o}}$ and $\varphi_{\mathrm{o}}^{(1)}$ can be tuned (the `active' dials), but all higher-order AD coefficients are determined by $\varphi_{\mathrm{o}}$ and $\varphi_{\mathrm{o}}^{(1)}$, and cannot be tuned independently (the `inactive' or `locked' dials).

We plot $\omega_{\mathrm{o}}\varphi_{\mathrm{o}}^{(1)}$ in Fig.~\ref{Fig:GratingCharacteristics}(a) as a function of $\alpha$ and $\varphi_{\mathrm{o}}$. Over a large portion of the parameter space $|\omega_{\mathrm{o}}\varphi_{\mathrm{o}}^{(1)}|\!\sim\!1$, and in general it increases only at large values of $\alpha$ and $\varphi_{\mathrm{o}}$. The values of the phase and group velocities depend on the observation axis selected. Along the $z$-axis (or normal to the grating), the phase velocity is $v_{\mathrm{ph}}\!=\!c/\cos{\varphi_{\mathrm{o}}}$, the first-order dispersion coefficients are $ck_{x}^{(1)}\!=\!\sin{\alpha}$, $ck_{z}^{(1)}\!=\!\tfrac{1-\sin{\alpha}\sin{\varphi_{\mathrm{o}}}}{\cos{\varphi_{\mathrm{o}}}}$, and the group velocity is [Fig.~\ref{Fig:GratingCharacteristics}(b)]:
\begin{equation}
\widetilde{v}=\frac{c\cos{\varphi_{\mathrm{o}}}}{1-\sin{\alpha}\sin{\varphi_{\mathrm{o}}}},
\end{equation}
which can be subluminal or superluminal, but is always positive. The group velocity along the $z$-axis is luminal $\widetilde{v}\!=\!c$ in two cases: (1) when $\varphi_{\mathrm{o}}\!=\!0^{\circ}$, or (2) when $\sin{\varphi_{\mathrm{o}}}\!=\!\tfrac{2\sin{\alpha}}{1+\sin^{2}{\alpha}}$. The regime enclosed between these two limits is superluminal $\widetilde{v}\!>\!c$, and that outside them is subluminal $\widetilde{v}\!<\!c$.

Significant deviation in $\widetilde{v}$ from $c$ along the $z$-axis requires large $\varphi_{\mathrm{o}}$. Changing the observation axis can have a profound effect on $\widetilde{v}$: $ck_{\bot}^{(1)}\!=\!\cos{\gamma}-\omega_{\mathrm{o}}\varphi_{\mathrm{o}}^{(1)}\sin{\gamma}$ along an observation axis making an angle $\gamma$ with the vector $\vec{k}_{\mathrm{o}}$. For example, negative values of $\widetilde{v}$ become accessible when $\tan{\delta_{\mathrm{o}}^{(1)}}\tan{\gamma}\!>\!1$.

The second-order dispersion coefficients along $z$ are:
\begin{equation}
c\omega_{\mathrm{o}}k_{x}^{(2)}=0,\,\,\,\, c\omega_{\mathrm{o}}k_{z}^{(2)}=-\frac{1}{\cos{\varphi_{\mathrm{o}}}}(\omega_{\mathrm{o}}\varphi_{\mathrm{o}}^{(1)})^{2};
\end{equation}
so that the GVD along $z$ is anomalous $k_{z}^{(2)}\!<\!0$ [Fig.~\ref{Fig:GratingCharacteristics}(c)]. Interestingly, we have $\delta_{\mathrm{o}}^{(2)}\!=\!-\varphi_{\mathrm{o}}$, so that $\vec{k}_{\mathrm{o}}^{(2)}$ is always aligned with the $z$-axis. Because $k_{x}^{(2)}\!=\!0$, the GVD remains anomalous along any observation axis. Therefore, a TPF produced by a grating \textit{always} experiences anomalous GVD.

Two configurations have been commonly utilized for GVD cancellation with a grating. The \textit{first} is normal incidence on the grating $\alpha\!=\!0$. As a result, the diffracted field is subluminal $\widetilde{v}\!=\!c\cos{\varphi_{\mathrm{o}}}$ along the $z$-axis. Moreover, $\delta_{\mathrm{o}}^{(1)}\!=\!\delta_{\mathrm{o}}^{(2)}\!=\!-\varphi_{\mathrm{o}}$, so that $\vec{k}_{\mathrm{o}}^{(1)}$ and $\vec{k}_{\mathrm{o}}^{(2)}$ coincide and are both normal to the grating, and the pulse front coincides with the plane of constant chirp. The anomalous GVD is tunable by changing $\varphi_{\mathrm{o}}$. The \textit{second} case is $\varphi_{\mathrm{o}}\!=\!0$, which results in $\widetilde{v}\!=\!c$ along the $z$-axis. Moreover, $\tan{\delta_{\mathrm{o}}^{(1)}}\!=\!\sin{\alpha}$, $c\omega_{\mathrm{o}}k_{z}^{(2)}\!=\!-\sin^{2}{\alpha}$, and $\delta_{\mathrm{o}}^{(2)}\!=\!0$, so that $\vec{k}_{\mathrm{o}}^{(2)}$ is again aligned with $\vec{k}_{\mathrm{o}}$. The anomalous axial GVD is tunable by changing $\alpha$.

In summary: (1) the AD-induced dispersion coefficients of all orders for a grating are determined by only two parameters (the incident and diffraction angles); (2) although arbitrary $\widetilde{v}$ can -- in principle -- be realized, it requires large propagation angles; and (3) only anomalous GVD is produced.

\subsection{AD produced by metasurfaces}

Dielectric metasurfaces have emerged as a tool to produce AD over a broad range of optical frequencies \cite{Arbabi2017Optica,Chen2018NatNanotech,Kamali2018Review,McClung2020Light, Chen2020NatRevMat, Dorrah2022Science}. These metasurfaces introduce AD into a normally incident optical field ($\alpha\!=\!0$), exerting control over $\varphi_{\mathrm{o}}^{(0)}$ and $\varphi_{\mathrm{o}}^{(1)}$. A single-surface meta-grating has been shown to produce a first-order AD coefficient in the range  $-0.6\leq\!\omega_{o}\varphi_{\mathrm{o}}^{(1)}\!\leq\!0.2$~rad at $\varphi_{\mathrm{o}}\!\approx\!10^{\circ}$ \cite{Arbabi2017Optica}, which is $3\times$ of AD produced by a conventional grating at \textit{normal} incidence ($\omega_{\mathrm{o}}\varphi_{\mathrm{o}}^{(1)}\!\approx\!-0.2$ from Eq.~\ref{Eq:ND_grating}) as well as inverting the AD sign by achieving positive $\varphi_{\mathrm{o}}^{(1)}$. Even more AD has been achieved by introducing a second surface to the meta-optic, resulting in $-1.5\leq\!\omega_{\mathrm{o}}\varphi_{\mathrm{o}}^{(1)}\!\leq\!0.5$~rad at $\varphi_{\mathrm{o}}\!\approx\!20^{\circ}$  \cite{McClung2020Light}. 
Although meta-gratings supply additional flexibility stemming from the frequency-dependent phase response of sub-wavelength structures, their demonstrations so far manifested independent control over only the first two AD coefficients $\varphi_{o}$ and $\varphi_{o}^{(1)}$ -- similarly to conventional gratings \cite{Arbabi2017Optica,Kamali2018Review}. In other words,  frequency-dependent response of meta-gratings restores the second degree of freedom lost while \textit{fixing} the angle of incidence $\alpha$, without offering any control over higher order AD coefficients. Moreover, a similar amount of $\varphi_{o}^{(1)}$ and the inverted sign of it demonstrated by these metasurfaces can be accomplished by tuning $\varphi_{o}$ or the angle of incidence $\alpha$. Therefore, independent control over higher-order AD terms $\varphi_{o}^{(n)}$ for $n\geq2$ is yet to be demonstrated using single-surface and double-surface meta-optics \cite{Arbabi2017Optica,McClung2020Light}.



More recent efforts in dispersion-engineered metasurfaces have been motivated by achromatic metalenses capable of focusing a broadband optical field in the visible and infrared spectrum \cite{Khorasaninejad2016Science, Wang2018Science, Li2021SciAdv}.  Achromatic focusing has been shown to require the phase delay and the group delay introduced by the meta-lens to be equal over its entire surface, while the group delay dispersion is kept zero \cite{Chen2018NatNanotech, Chen2020NatRevMat}. Therefore, the dispersion-engineering methodologies developed for achromatic meta-surfaces \cite{Chen23NC, Hu23NC}, in principle, can be extended for independent control over AD coefficients beyond the capabilities of conventional gratings, which is yet to be demonstrated. 

\section{Arbitrary angular-dispersion control}\label{Section:ADControl}

\subsection{Structure of the relationship between dispersion and AD}

Typical optical components offer control over only a few AD orders. Indeed, until our recent work on STWPs \cite{Hall21OLNormalGVD}, no TPF has realized normal GVD in free space. We pose the following question: What opportunities does the tunability of multiple AD orders provide in optics?

\begin{figure}[t!]
\centering
\includegraphics[width=8.6cm]{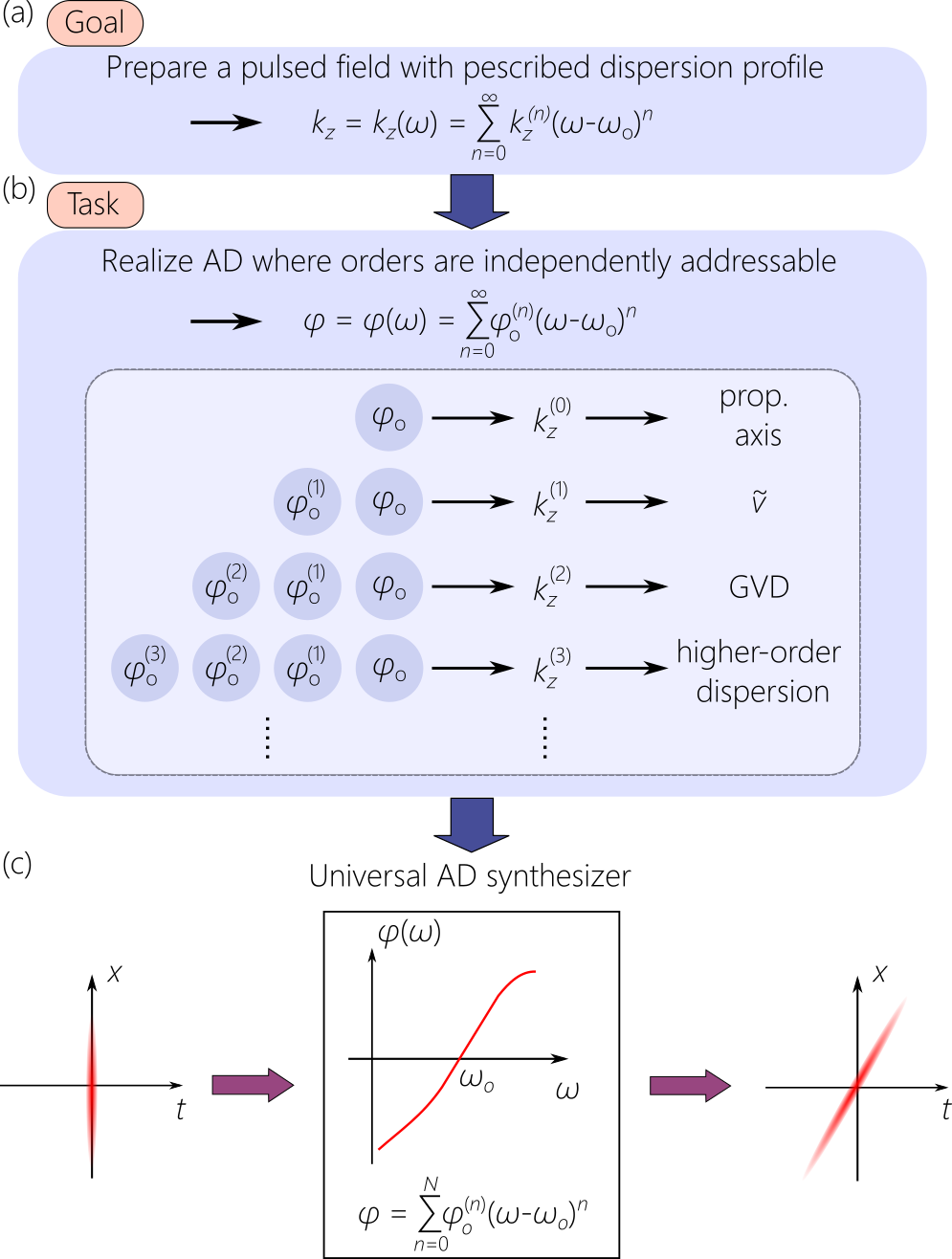}
\caption{Algorithmic procedure for synthesizing an arbitrary dispersion profile via AD. (a) The \textit{goal} is to produce a pulsed optical field with an arbitrary dispersion profile in which all the dispersion coefficients $\{k_{z}^{(n)}\}$ are prescribed. (b) Because of the hierarchical structure of the relationship between $\{k_{z}^{(n)}\}$ and $\{\varphi_{\mathrm{o}}^{(n)}\}$, achieving this goal implies a challenging \textit{task}: introducing AD whose coefficients of all orders $\{\varphi_{\mathrm{o}}^{(n)}\}$ are independently tunable. (c) Realizing this task necessitates constructing a universal AD synthesizer.}
\label{Fig:Algorithmic}
\end{figure}

According to the perturbative theory of AD in Section~\ref{section:Theory of conventional angular dispersion}, the dispersion coefficients $\{k_{z}^{(n)}\}$ form a hierarchical structure with respect to the AD coefficients $\{\varphi_{\mathrm{o}}^{(n)}\}$; i.e., $k_{z}^{(n)}$ depends on all the AD coefficients $\{\varphi_{\mathrm{o}}^{(j)}\}_{j=0}^{n}$ of order $j\!\leq\!n$. Consequently, synthesizing an optical field with prescribed dispersion profile $k_{z}(\omega_{\mathrm{o}}+\Omega)\!=\!\sum_{n=0}^{\infty}\tfrac{1}{n!}k_{z}^{(n)}\Omega^{n}$ necessitates mapping the \textit{target} set of dispersion coefficients $\{k_{z}^{(n)}\}$ to a \textit{required} set of AD coefficients $\{\varphi_{\mathrm{o}}^{(n)}\}$, as illustrated in Fig.~\ref{Fig:Algorithmic}. In other words, full control over the dispersion profile necessitates independent tunability of all the relevant AD coefficients. 

This cannot be achieved using conventional optical components, which are typically characterized by a small number of physical parameters that render only a \textit{subset} of coefficients $\{\varphi_{\mathrm{o}}^{(n)}\}$ accessible. Instead, a universal AD synthesizer, which is an optical arrangement having a large number of independent physical degrees of freedom, is needed to provide full control over $\varphi(\omega)$. We discuss next the capabilities made possible as control over an increasing number of degrees-of-freedom is made available to a universal AD synthesizer.

\subsection{Control over first-order AD}

\begin{figure}[t!]
\centering
\includegraphics[width=8.6cm]{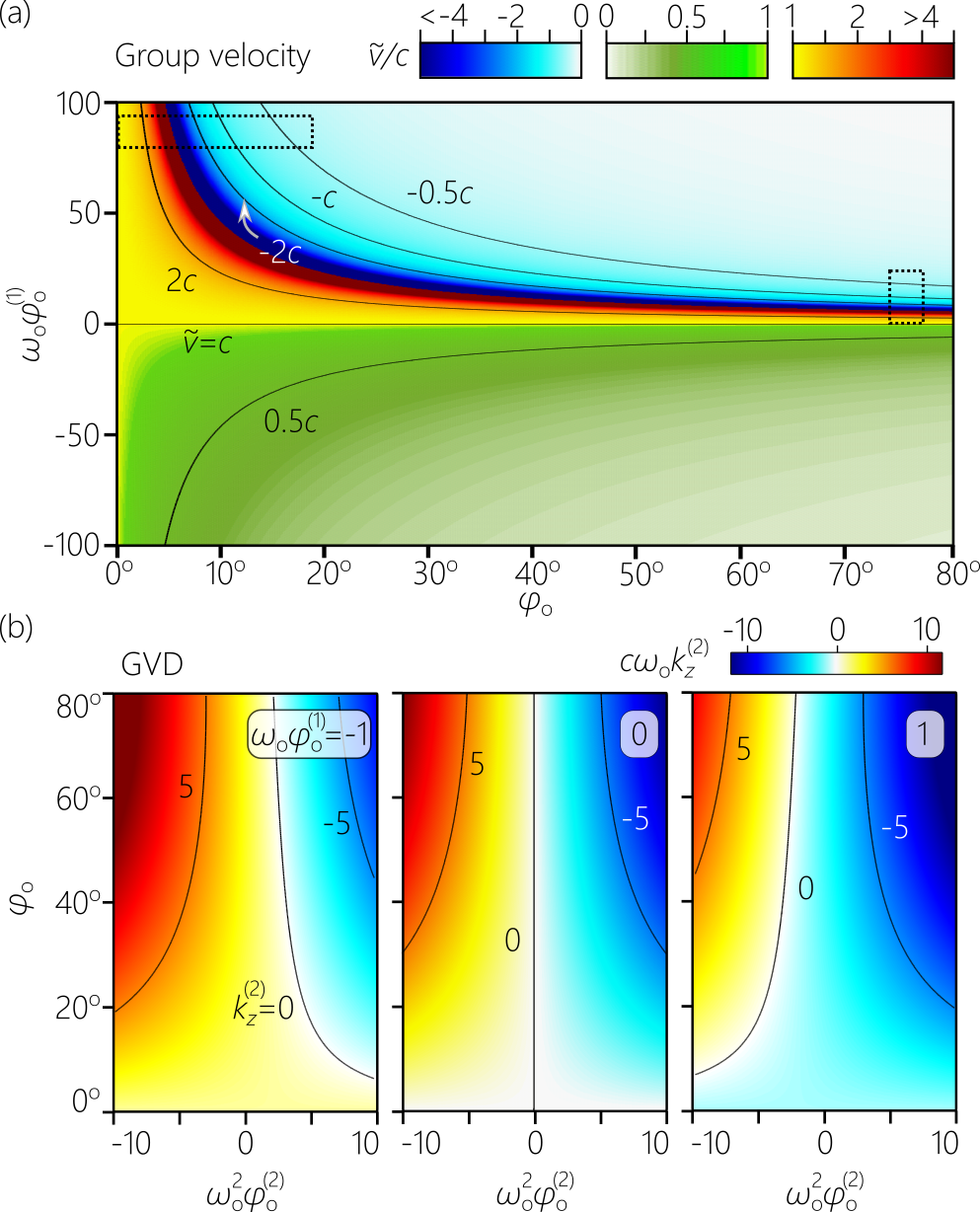}
\caption{(a) Group velocity $\widetilde{v}$ along the $z$-axis (Eq.~\ref{Eq:GroupVelocityAngularDispersion}) as a function of $\varphi_{\mathrm{o}}$ and $\omega_{\mathrm{o}}\varphi_{\mathrm{o}}^{(1)}$. We employ three color palettes: green for the subluminal regime $0\!<\!\widetilde{v}\!<\!c$, red for the superluminal $c\!<\!\widetilde{v}\!<\!\infty$, and blue for negative values $\widetilde{v}\!<\!0$. The black curves are contours of constant $\widetilde{v}$. The dashed box on the right identifies a region of large $\varphi_{\mathrm{o}}$ in which small changes in $\omega_{\mathrm{o}}\varphi_{\mathrm{o}}^{(1)}$ result in large changes in $\widetilde{v}$. The role of $\varphi_{\mathrm{o}}$ and $\omega_{\mathrm{o}}\varphi_{\mathrm{o}}^{(1)}$ are reversed in the dashed box in the top-left corner. (b) Plots of the GVD coefficient as a function of $\omega_{\mathrm{o}}^{2}\varphi_{\mathrm{o}}^{2}$ and $\varphi_{\mathrm{o}}$, with fixed  $\omega_{\mathrm{o}}\varphi_{\mathrm{o}}^{(1)} = -1$, $0$, and $1$ in each panel. The black curves are contours of constant $k_{z}^{(2)}$.}
\label{Fig:GeneralGroupVelocity}
\end{figure}

Independent control over the AD coefficients $\varphi_{\mathrm{o}}$ and $\varphi_{\mathrm{o}}^{(1)}$ helps realize any value of $\widetilde{v}$ in principle, which is brought out in the plot of $\widetilde{v}$ (Eq.~\ref{Eq:GroupVelocityAngularDispersion}) in Fig.~\ref{Fig:GeneralGroupVelocity}(a). We identify the subluminal, superluminal, and negative-$\widetilde{v}$ regimes with different color palettes. The plot in Fig.~\ref{Fig:GeneralGroupVelocity}(a) therefore identifies the angular \textit{resources} needed to realize any prescribed $\widetilde{v}$. There is a continuum of pairs of values for $\varphi_{\mathrm{o}}$ and $\omega_{\mathrm{o}}\varphi_{\mathrm{o}}^{(1)}$ that yields any desired $\widetilde{v}\!=\!c/\widetilde{n}$ along the 1D parametric curve: $\tfrac{\cos{\varphi_{\mathrm{o}}}-\widetilde{n}}{\sin{\varphi_{\mathrm{o}}}}\!=\!\omega_{\mathrm{o}}\varphi_{\mathrm{o}}^{(1)}$.

To produce a wave packet whose group velocity deviates strongly from $c$ in free space, one needs an optical system that provides one of the following two options:
\begin{enumerate}
\item A large $\varphi_{\mathrm{o}}$ and modest $\omega_{\mathrm{o}}\varphi_{\mathrm{o}}^{(1)}$, corresponding to the highlighted region to the right in Fig.~\ref{Fig:GeneralGroupVelocity}(a). Here, small changes in $\omega_{\mathrm{o}}\varphi_{\mathrm{o}}^{(1)}$ at large $\varphi_{\mathrm{o}}$ can rapidly tune $\widetilde{v}$.
\item A large $\omega_{\mathrm{o}}\varphi_{\mathrm{o}}^{(1)}$ at modest $\varphi_{\mathrm{o}}$, corresponding to the highlighted region in the top-left corner of Fig.~\ref{Fig:GeneralGroupVelocity}(a). Here, small changes in $\varphi_{\mathrm{o}}$ at large $\omega_{\mathrm{o}}\varphi_{\mathrm{o}}^{(1)}$ can rapidly tune $\widetilde{v}$.
\end{enumerate}

Figure~\ref{Fig:GeneralGroupVelocity}(a) highlights the need for independent control over $\varphi_{\mathrm{o}}$ and $\varphi_{\mathrm{o}}^{(1)}$ to tune $\widetilde{v}$, However, such control is not always available. For example, a grating does not provide the full span of values for $\widetilde{v}$ along the normal to the grating [Fig.~\ref{Fig:GratingCharacteristics}(b)]. Another example is X-waves where $\varphi_{\mathrm{o}}^{(1)}\!=\!0$ and $\widetilde{v}\!=\!\tfrac{c}{\cos{\varphi_{o}}}$ \cite{Saari97PRL}, which depends solely on $\varphi_{\mathrm{o}}$. Because $\varphi_{\mathrm{o}}^{(1)}=0$, $\widetilde{v}$ deviates significantly from $c$ only at large $\varphi_{\mathrm{o}}$ in the non-paraxial regime \cite{Yessenov22AOP}, corresponding to the dashed box on the right of Fig.~\ref{Fig:GeneralGroupVelocity}(a). In the paraxial regime (small $\varphi_{\mathrm{o}}$), only minute deviation from $c$ is achievable; e.g., $\widetilde{v}\!\approx\!1.004c$ at $\varphi_{\mathrm{o}}\!=\!5^{\circ}$. In principle, one can produce a paraxial TPF whose group velocity deviates significantly from $c$ in free space by increasing $\varphi_{\mathrm{o}}^{(1)}$ rather than increasing $\varphi_{\mathrm{o}}$. However, besides the inherent difficulty in producing large values of $\varphi_{\mathrm{o}}^{(1)}$, this in turn increases the GVD accrued by the TPF, thus limiting the propagation distance over which $\widetilde{v}$ can be reliably measured. Consequently, the reported group velocities of TPFs and X-waves to date have remained close to $c$ ($1.00022c$ in \cite{Bonaretti09OE}, $1.00012c$ in \cite{Bowlan09OL}, and $1.00015c$ in \cite{Kuntz09PRA}), and $0.99999c$ for a pulsed Bessel beam \cite{Giovannini15Science}.

\subsection{Control over second-order AD}

As shown in Eq.~\ref{Eq:SecondOrderDispersionCoefficients}, the GVD incurred in free space by on-axis optical fields incorporating AD is always anomalous. Off-axis propagation ($\varphi_{\mathrm{o}}\neq0$) allows for AD-induced normal GVD in free space:
\begin{equation}\label{eq:OffaxisGVD}
c\omega_{\mathrm{o}}k_{z}^{(2)}=-(\omega_{\mathrm{o}}\varphi_{\mathrm{o}}^{(1)})^{2}\cos{\varphi_{\mathrm{o}}}-(\omega_{\mathrm{o}}^{2}\varphi_{\mathrm{o}}^{(2)}+2\omega_{\mathrm{o}}\varphi_{\mathrm{o}}^{(1)})\sin{\varphi_{\mathrm{o}}}.
\end{equation}
Tuning $\varphi_{\mathrm{o}}$, $\varphi_{\mathrm{o}}^{(1)}$, and $\varphi_{\mathrm{o}}^{(2)}$ independently helps span both normal and anomalous GVD [Fig.~\ref{Fig:GeneralGroupVelocity}(b)]. For example, in the special case proposed in \cite{Porras03PRE2}, $\varphi_{\mathrm{o}}^{(1)}\!=\!0$, so that $c\omega_{\mathrm{o}}k_{z}^{(2)}\!=\!-\omega_{\mathrm{o}}^{2}\varphi_{\mathrm{o}}^{(2)}\sin{\varphi_{\mathrm{o}}}$, whereupon normal GVD is produced by ensuring that $\varphi_{\mathrm{o}}$ and $\varphi_{\mathrm{o}}^{(2)}$ have opposite signs. More generally, without requiring that $\varphi_{\mathrm{o}}^{(1)}\!=\!0$, normal GVD can be realized when $\tan{\delta_{\mathrm{o}}^{(2)}}\tan{\psi}\!>\!1$, so that $\cos{(\delta_{\mathrm{o}}^{(2)}+\varphi_{\mathrm{o}})}\!<\!0$ or $\delta_{\mathrm{o}}^{(2)}+\varphi_{\mathrm{o}}\!>\!90^{\circ}$. Nevertheless, independent control over $\varphi_{\mathrm{o}}^{(2)}$ is a capability \textit{not} provided by any known optical device. 

\begin{figure*}[t!]
\centering
\includegraphics[width=17.6cm]{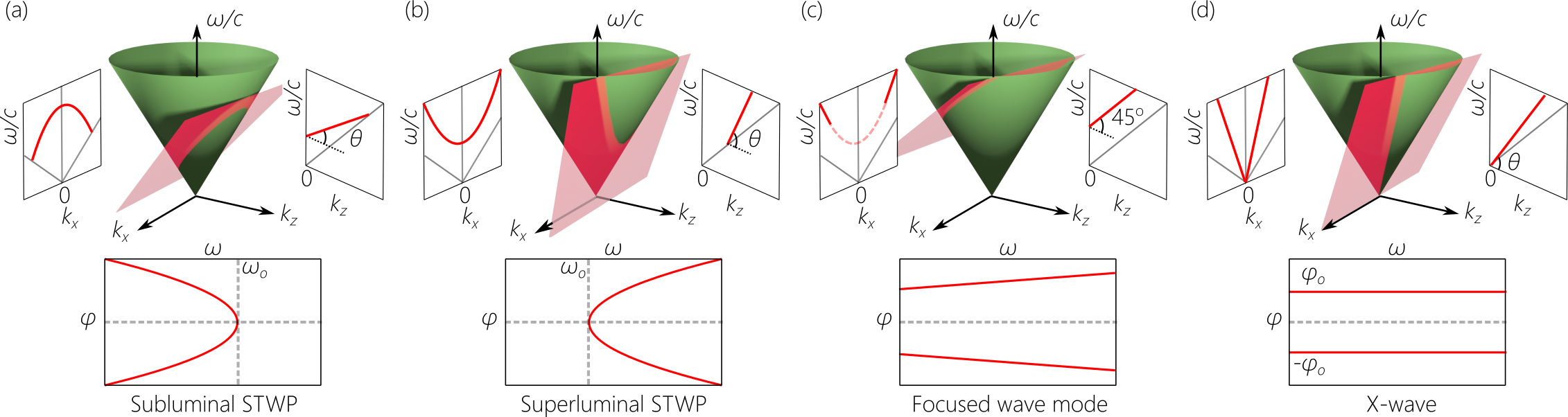}
\caption{Spectral support for propagation-invariant STWPs on the surface of the free-space light-cone and their associated AD profile $\varphi(\omega)$. (a) Subluminal ($\theta\!<\!45^{\circ}$) baseband STWP with $k_{z}\!=\!k_{\mathrm{o}}$ at $k_{x}\!=\!0$; (b) superluminal ($\theta\!>\!45^{\circ}$) baseband STWP with $k_{z}\!=\!k_{\mathrm{o}}$ at $k_{x}\!=\!0$; (c) luminal ($\theta\!=\!45^{\circ}$) FWM with $k_{z}\!=\!-k_{\mathrm{o}}$ at $k_{x}\!=\!0$; and (d) superluminal ($\theta\!>\!45^{\circ}$) X-wave with $k_{z}\!=\!0$ at $k_{x}\!=\!0$.}
\label{Fig:LightConesForSTWavePackets}
\end{figure*}

\section{What is non-differentiable angular dispersion?}\label{Section:NonDifferentiableAD}

The constraints stemming from the differentiability of the AD as outlined in the Introduction have been permanent features of the optics landscape. What exactly is \textit{non-differentiable AD}, and can it be readily introduced into generic optical fields? What are the novel opportunities that are made possible by non-differentiable AD?

It is useful to first state what non-differentiable AD is \textit{not}: it does not refer to pathological or exotic mathematical functions for $\varphi(\omega)$. Rather, non-differentiable AD can be implemented while $\varphi(\omega)$ remains finite, continuous, and differentiable everywhere except at one (or more) frequency $\omega_{\mathrm{o}}$. This frequency becomes a natural terminus for the spectrum, and is denoted the non-differentiable frequency [Fig.~\ref{Fig:AngularDispersion}(d)]. For example, consider an optical field incorporating AD where $\varphi(\omega)\!\propto\!\sqrt{\omega-\omega_{\mathrm{o}}}$, so that $\omega_{\mathrm{o}}$ is the lower limit on the spectrum. The function $\sqrt{x}$ is not differentiable at $x\!=\!0$, and such an optical field is thus endowed with \textit{non-differentiable AD}, and $\omega\!=\!\omega_{\mathrm{o}}$ is its non-differentiable frequency.

No individual optical component can produce to date such a field structure. However, non-differentiable AD is necessary for a broad swath of desirable features that are inaccessible to conventional optical fields, including:
\begin{enumerate}
\item Tuning the \textit{on-axis} group velocity in free space away from $c$.
\item Realizing propagation-\textit{invariant} wave packets endowed with AD; i.e., a pulsed beam that incorporates AD and yet is free of dispersion to all orders. 
\item Tuning the magnitude and sign of a single dispersion coefficient while eliminating all others in an on-axis field. For example, realizing normal GVD in free space when $\varphi_{\mathrm{o}}=0$.
\end{enumerate}
We emphasize that these three tasks are \textit{impossible} to achieve when one is restricted to differentiable AD.

\subsection{Tuning the on-axis group velocity via non-differentiable AD}

Can the group velocity of a wave packet deviate from $c$ along its propagation axis ($\varphi_{\mathrm{o}}\!=\!0$) in presence of AD? According to Eq.~\ref{Eq:GroupVelocityAngularDispersion}, $\widetilde{v}\!=\!c$ along $\vec{k}_{\mathrm{o}}$ because $\varphi\!\rightarrow\!0$ and $\omega\tfrac{d\varphi}{d\omega}\sin{\varphi}\!\approx\!\omega\varphi\tfrac{d\varphi}{d\omega}\!\rightarrow\!0$ when $\omega\!\rightarrow\!\omega_{\mathrm{o}}$. However, if $\varphi(\omega)$ is \textit{not} differentiable at $\omega_{\mathrm{o}}$, then $\omega\varphi\tfrac{d\varphi}{d\omega}$ may take on a \textit{finite} value when $\omega\!\rightarrow\!\omega_{\mathrm{o}}$. For example, when $\varphi(\omega)\!=\!\varphi(\omega_{\mathrm{o}}+\Omega)\!=\!\sqrt{2\eta\tfrac{\Omega}{\omega_{\mathrm{o}}}}$, which is \textit{not differentiable} at $\Omega\!=\!0$ ($\eta$ is a dimensionless, positive constant), then $\omega\varphi\tfrac{d\varphi}{d\omega}\!\rightarrow\!\eta$ when $\omega\!\rightarrow\!\omega_{\mathrm{o}}$, in which case $\widetilde{v}\!=\!\tfrac{c}{1-\eta}$. Therefore, non-differentiable AD can help tune $\widetilde{v}$ along $\vec{k}_{\mathrm{o}}$ in the paraxial regime by simply adjusting $\eta$. Indeed, any paraxial pulsed field in which $\widetilde{v}$ deviates significantly from $c$ in free space will be endowed at least approximately with non-differentiable AD. For example, the `flying focus' in \cite{Froula18NP} has a widely tunable $\widetilde{v}$, and we thus expect that some form of non-differentiable AD underpins this field structure. In contrast, the `achromatic flying focus' in \cite{Pigeon24OE} is endowed with differentiable AD, resulting in a dramatically reduced span of values for $\widetilde{v}$.

\subsection{Non-differentiable AD can lead to propagation invariance}

A propagation-invariant wave packet is one that is transported rigidly in free space without diffraction or dispersion at a fixed group velocity $\widetilde{v}\!=\!c/\widetilde{n}$ \cite{Reivelt03arxiv,Turunen10PO,FigueroaBook14}. This entails the absence of axial dispersion of all orders: $k_{z}^{(n)}\!=\!0$ for $n\!\geq\!2$. Consequently,  the group index must be \textit{frequency-independent}:
\begin{equation}
\widetilde{n}(\omega)=ck_{z}^{(1)}=\tfrac{d}{d\omega}\{\omega\cos{\varphi}\}=\widetilde{n}.
\end{equation}
Integration yields $k_{z}\!=\!k\cos{\varphi}\!=\!\tfrac{\Omega}{\widetilde{v}}+k_{\mathrm{a}}$, where $\Omega\!=\!\omega-\omega_{\mathrm{o}}$ and $k_{\mathrm{a}}$ is a constant. Three different scenarios can lead to propagation invariance.

\textbf{1. Baseband STWPs.} By requiring that $\omega\!=\!\omega_{\mathrm{o}}$ propagate along $z$ ($\varphi_{\mathrm{o}}\!=\!0^{\circ}$ and $k_{\mathrm{a}}\!=\!k_{\mathrm{o}}$), then $k_{z}\!=\!k_{\mathrm{o}}+(\omega-\omega_{\mathrm{o}})/\widetilde{v}$, whereupon $\varphi(\Omega)\approx\sqrt{2(1-\widetilde{n})\tfrac{\Omega}{\omega_{\mathrm{o}}}}\!\propto\!\sqrt{\Omega}$, which is \textit{not} differentiable at $\Omega\!=\!0$. Here $\omega\!>\omega_{\mathrm{o}}$ when $\widetilde{n}\!<\!1$ (superluminal), and $\omega\!<\omega_{\mathrm{o}}$ when $\widetilde{n}\!>\!1$ (subluminal). This encompasses the family of `baseband' STWPs \cite{ZamboniRached2008PRA,Yessenov19PRA}, whose name refers to the fact that the spatial spectrum of such a wave packet is in the vicinity of $k_{x}\!=\!0$.  

Geometric intuition can be gleaned from examining the spectral support for propagation-invariant baseband STWP along the conic section at the intersection of the light-cone with the plane $k_{z}\!=\!k_{\mathrm{o}}+\Omega/\widetilde{v}$ [Fig.~\ref{Fig:LightConesForSTWavePackets}(a)], which makes an angle $\theta$ (the spectral tilt angle) with the $k_{z}$-axis, resulting in $\widetilde{v}\!=\!c\tan{\theta}$. The non-differentiable frequency $\omega_{\mathrm{o}}$ is the maximum when $0\!<\!\theta\!<\!45^{\circ}$ (subluminal regime, $\widetilde{v}\!<\!c$) [Fig.~\ref{Fig:LightConesForSTWavePackets}(a)], or the minimum when $45^{\circ}\!<\!\theta\!<\!180^{\circ}$ (superluminal regime, $\widetilde{v}\!>\!c$, and negative-$\widetilde{v}$ regime, $\widetilde{v}\!<\!0$) [Fig.~\ref{Fig:LightConesForSTWavePackets}(b)]. 

\textbf{2. Sideband STWPs.} A second solution exists when $k_{\mathrm{a}}\!=\!-k_{\mathrm{o}}$, and $k_{z}\!=\!-k_{\mathrm{o}}+(\omega-\omega_{\mathrm{o}})/\widetilde{v}$, which corresponds to the family of \textit{sideband} STWPs, including focus-wave modes (FWMs) discovered by Brittingham in 1983 \cite{Brittingham83JAP}, for which $\widetilde{v}\!=\!c$ and $k_{z}\!=\!-k_{\mathrm{o}}+\Omega/c$ \cite{Yessenov19PRA}. The `sideband' moniker refers to the exclusion of spatial frequencies in the vicinity of $k_{x}\!=\!0$ because they are associated with $k_{z}\!<\!0$, a regime that is not consistent with relativistic causality \cite{Heyman87JOSAA} (only $k_{z}\!>\!0$ is permissible). Here $\varphi(\omega)$ is also non-differentiable at $\omega\!=\!\omega_{\mathrm{o}}$, $\varphi(\Omega)\!=\!\pi-\sqrt{2(1+\widetilde{n})\tfrac{\Omega}{\omega_{\mathrm{o}}}}$. However, $\omega_{\mathrm{o}}$ occurs at $k_{z}(\omega_{\mathrm{o}})\!=\!-k_{\mathrm{o}}$, which is excluded on physical grounds \cite{Heyman87JOSAA,Yessenov19PRA,Yessenov22AOP}. Therefore, although the non-differentiable frequency is inaccessible physically, its existence is nevertheless necessary for the propagation invariance of sideband STWPs.

The spectral support for sideband STWPs is the conic section at the intersection of the light-cone with the plane $k_{z}\!=\!-k_{\mathrm{o}}+\Omega/\widetilde{v}$, which also makes an angle $\theta$ with respect to the $k_{z}$-axis, with $\widetilde{v}\!=\!c\tan{\theta}$ [Fig.~\ref{Fig:LightConesForSTWavePackets}(c)]. The point $(k_{x},k_{z},\tfrac{\omega}{c})\!=\!(0,-k_{\mathrm{o}},k_{\mathrm{o}})$ at which the AD is non-differentiable is not part of the spectrum because $k_{z}\!<\!0$ at this point.

\begin{figure*}[t!]
\centering
\includegraphics[width=17.6cm]{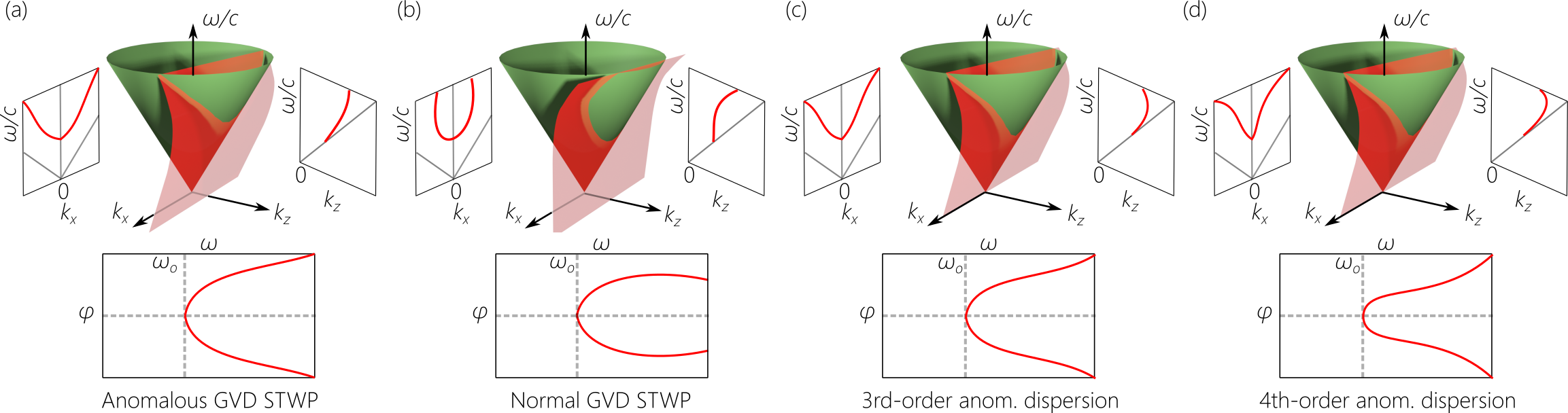}
\caption{Same as Fig.~\ref{Fig:LightConesForSTWavePackets} for dispersive STWPs. The spectral support is the intersection of the light-cone with a planar curved surface that is parallel to the $k_{x}$-axis rather than a plane. (a) STWP with anomalous GVD $k_{z}^{(2)}\!<\!0$; (b) STWP with normal GVD $k_{z}^{(2)}\!>\!0$; (c) STWP with anomalous third-order dispersion coefficient $k_{z}^{(3)}\!<\!0$; and (d) STWP with anomalous fourth-order dispersion coefficient $k_{z}^{(4)}\!<\!0$.}
\label{Fig:LightConesWithDispersion}
\end{figure*}

\textbf{3. X-waves.} Finally, maintaining $\widetilde{n}(\omega)\!=\!\widetilde{n}$ is possible when $\varphi(\omega)\!=\!\varphi_{\mathrm{o}}$, which corresponds to X-waves that are thus AD-free with $\widetilde{n}\!=\!\cos{\varphi_{\mathrm{o}}}$. Here $k_{z}\!=\!\tfrac{\omega}{c}\cos{\varphi_{\mathrm{o}}}$, so their superluminal phase and group velocities $v_{\mathrm{ph}}\!=\!\widetilde{v}\!=\!c/\cos{\varphi_{\mathrm{o}}}$ stem from a purely geometric origin \cite{Chiao02OPN}. The spectral support for AD-free X-waves is the pair of straight lines at the intersection of the light-cone with the plane $k_{z}\!=\!\tfrac{\omega}{c}\cos{\varphi_{\mathrm{o}}}$ that passes through the origin [Fig.~\ref{Fig:LightConesForSTWavePackets}(d)], where $\cos{\varphi_{\mathrm{o}}}\!=\!\cot{\theta}$. The spectral tilt angle $\theta$ is restricted to the superluminal range $45^{\circ}\!<\!\theta\!<\!90^{\circ}$.

It is instructive to observe that a similar concept applies to \textit{chromatic} dispersion. Elimination of all higher-order dispersion coefficients requires that the group index be independent of frequency: $n_{\mathrm{g}}\!=\!n+\omega\tfrac{dn}{d\omega}\!=\!\tfrac{d(\omega n)}{d\omega}$ is $\omega$-independent, so $n(\omega)\!=\!n_{\mathrm{g}}+\tfrac{\sigma}{\omega}$, where $\sigma\!=\!\omega_{\mathrm{o}}(n_{\mathrm{o}}-n_{\mathrm{g}})$ and $n(\omega_{\mathrm{o}})\!=\!n_{\mathrm{o}}$. The refractive index is $n(\omega)\!=\!n_{\mathrm{g}}+(n_{\mathrm{o}}-n_{\mathrm{g}})\tfrac{\omega_{\mathrm{o}}}{\omega}$, which entails that $\tfrac{dk}{d\omega}\!=\!\tfrac{n_{\mathrm{g}}}{c}\!=\!\tfrac{1}{v_{\mathrm{g}}}$ and $\tfrac{d^{n}k}{d\omega^{n}}\!=\!0$ for $n\!\geq\!2$. However, no known optical material displays this behavior, at least away from a resonance.

\subsection{Tuning a single dispersion order via non-differentiable AD} \label{Section: controllingSingleDispersion_via_NDAD}

Another unique feature enabled by non-differentiable AD is control over both magnitude \textit{and} sign of any particular order of axial dispersion $k_{z}^{(n)}$, $n\!\geq\!2$, while simultaneously eliminating all other orders, $k_{z}^{(m)}\!=\!0$ for $m\!\neq\!n$. As discussed in Section~\ref{Section:ADControl}, this is impossible with differentiable AD.

Consider the example of tuning the GVD coefficient $k_{z}^{(2)}$ in an on-axis field while maintaining $k_{z}^{(m)}\!=\!0$ for $m\!\geq\!3$, which requires that $k_{z}^{(2)}(\omega)\!=\!k_{2}$ be a frequency-independent constant. Writing $c\omega k_{z}^{(2)}\!=\!\tfrac{d}{d\omega}\{\omega^{2}\tfrac{d\cos{\varphi}}{d\omega}\}$ (Eq.~\ref{eq:OffaxisGVD}), and integrating twice yields $k_{z}(\omega)\!=\!k\cos{\varphi(\omega)}\!=\!\tfrac{1}{2}k_{2}\omega^{2}+\tfrac{1}{c}\sigma_{1}+\tfrac{\omega}{c}\sigma_{2}$; where $\sigma_{1}$ and $\sigma_{2}$ are constants. For baseband STWPs where $\varphi_{\mathrm{o}}\!=\!0$, $k_{z}(\omega_{\mathrm{o}})\!=\!k_{\mathrm{o}}$, and $\tfrac{dk_{z}}{d\omega}\big|_{\omega_{\mathrm{o}}}=\tfrac{1}{\widetilde{v}}$, we have $k_{z}\!=\!k_{\mathrm{o}}+\tfrac{1}{\widetilde{v}}\Omega+\tfrac{1}{2}k_{2}\Omega^{2}$. Crucially, the truncation of this dispersion relationship at second order is \textit{not} an approximation. The wave packet has a group velocity $\widetilde{v}$, GVD coefficient $k_{2}$, and \textit{all} higher-order dispersion coefficients are eliminated. Once again, the AD is non-differentiable at $\omega\!\rightarrow\!\omega_{\mathrm{o}}$. Two unique features emerge here: (1) this result is agnostic with respect to the sign of $k_{2}$, so that normal or anomalous GVD can be produced along the optical axis, in contradiction with the MGF theorem; and (2) \textit{all} higher-order dispersion coefficients are eliminated. In contrast, an on-axis TPF can experience only anomalous GVD, and all higher-order dispersion terms persist, although they generally diminish with increasing order.

The spectral support for dispersive STWPs is the intersection of the light-cone with \textit{planar curved surfaces} [Fig.~\ref{Fig:LightConesWithDispersion}] rather than \textit{planes} [Fig.~\ref{Fig:LightConesForSTWavePackets}]. These planar curved surfaces are parallel to the $k_{x}$-axis, and their projection onto the $(k_{z},\tfrac{\omega}{c})$-plane is a curve corresponding to the target axial-dispersion spectral profile. Critically, the 1D spectral trajectory has a maximum (when $\widetilde{v}\!<\!c$) or minimum (when $\widetilde{v}\!>\!c$) at the point $(k_{x},k_{z},\tfrac{\omega}{c})\!=\!(0,k_{\mathrm{o}},k_{\mathrm{o}})$, at which the AD is non-differentiable. When the GVD is normal ($k_{z}^{(2)}\!>\!0$), the surface is curved away from the light-line as shown in Fig.~\ref{Fig:LightConesWithDispersion}(a). In contrast, anomalous GVD ($k_{z}^{(2)}\!<\!0$) requires a surface that curves back towards the light-line as shown in Fig.~\ref{Fig:LightConesWithDispersion}(b). Therefore, normal GVD requires larger propagation angles than a propagation-invariant STWP having the same group velocity, whereas anomalous GVD requires smaller angles. Because of this difference, there are distinct limits on the achievable combined values of $\widetilde{v}$ and $k_{z}^{(2)}$; see Ref.~\cite{Yessenov21ACSP}. 

The same approach can be followed for any other dispersion order selected so that $k_{z}\!=\!k_{\mathrm{o}}+\Omega/\widetilde{v}+\tfrac{1}{n!}k_{z}^{(n)}\Omega^{n}$, as shown in Fig.~\ref{Fig:LightConesWithDispersion}(c) for $n=3$, and Fig.~\ref{Fig:LightConesWithDispersion}(d) for $n=4$. This approach can be easily generalized to arbitrary superpositions of dispersion terms. The structure of the dispersion profile and the requisite curved surface becomes more complex, but the general features of the spectral support are similar \cite{Yessenov21ACSP}. 

\subsection{Pulse-front tilt in presence of non-differentiable AD}

\begin{figure}[t!]
\centering
\includegraphics[width=8.6cm]{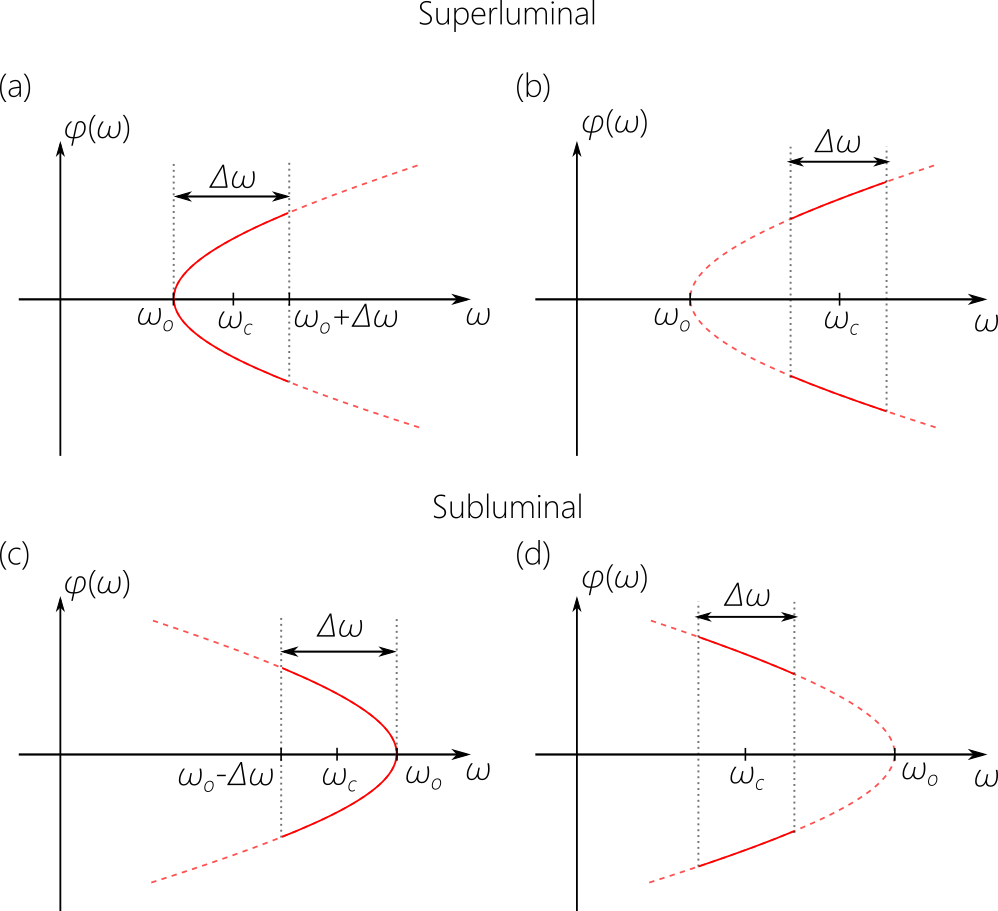}
\caption{Impact of the proximity of the STWP-spectrum to the non-differentiable frequency $\omega\!=\!\omega_{\mathrm{o}}$ on the bandwidth-dependence of its pulse-front tilt angle $\delta_{\mathrm{o}}^{(1)}$. (a) The AD profile $\varphi(\omega)$ for a superluminal propagation-invariant STWP whose spectrum includes the non-differentiable frequency $\omega_{\mathrm{o}}$, and (b) one that does \textit{not} include $\omega_{\mathrm{o}}$. (c,d) Same as (a,b) for a subluminal propagation-invariant STWP.}
\label{Fig:Parabola}
\end{figure}

We have shown that $\omega_{\mathrm{o}}$ is a natural spectral endpoint for baseband STWPs [Fig.~\ref{Fig:Parabola}]. If the bandwidth is $\Delta\omega$, then the lowest spectral range extends from $\omega_{\mathrm{o}}$ to $\omega_{\mathrm{o}}+\Delta\omega$ for superluminal STWPs [Fig.~\ref{Fig:Parabola}(a)]. Because $\varphi_{\mathrm{o}}^{(1)}$ is not defined at $\omega_{\mathrm{o}}$, we cannot use Eq.~\ref{Eq:AngleOfPulseFrontTilt} to evaluate the pulse-front tilt angle $\delta_{\mathrm{o}}^{(1)}$. However, we have shown in \cite{Hall21OL} that the ansatz
\begin{equation}\label{Eq:SpectralTiltAngleSTWP}
\tan{\delta_{\mathrm{o}}^{(1)}}=\sqrt{\frac{|1-\widetilde{n}|}{\Delta\omega/\omega_{\mathrm{o}}}}\propto\frac{1}{\sqrt{\Delta\omega}}
\end{equation}
fits measurements of the pulse-front tilt, thereby indicating that $\delta_{\mathrm{o}}^{(1)}$ depends inversely on the square-root of the bandwidth, in violation of the accepted universality of Eq.~\ref{Eq:AngleOfPulseFrontTilt} whereby $\delta_{\mathrm{o}}^{(1)}$ is bandwidth-independent.

A useful perspective arises from taking the carrier frequency to be the mid-spectral range frequency $\omega_{\mathrm{c}}\!=\!\omega_{\mathrm{o}}+\Delta\omega/2$ where the AD is differentiable, rather than $\omega_{\mathrm{o}}$ where the AD is non-differentiable. The spectrum now extends over the range $\omega_{\mathrm{c}}-\tfrac{1}{2}\Delta\omega$ to $\omega_{\mathrm{c}}+\tfrac{1}{2}\Delta\omega$. At $\omega\!=\!\omega_{\mathrm{c}}$ we have:
\begin{equation}\label{Eq:SpectralTiltAngleSTWPAwayFromNDPoint}
\tan{\delta_{\mathrm{c}}^{(1)}}=\omega_{\mathrm{c}}\varphi_{\mathrm{c}}^{(1)}=\frac{\omega_{\mathrm{c}}}{\omega_{\mathrm{o}}}\sqrt{\frac{(1-\widetilde{n})/2}{\frac{\omega_{\mathrm{c}}}{\omega_{\mathrm{o}}}-1}}.
\end{equation}
When $\omega_{\mathrm{c}}$ is close to $\omega_{\mathrm{o}}$ and $\Delta\omega\!=\!2(\omega_{\mathrm{c}}-\omega_{\mathrm{o}})\!\ll\!\omega_{\mathrm{c}},\omega_{\mathrm{o}}$, then $\tan{\delta_{\mathrm{c}}^{(1)}}\!\approx\!\sqrt{\tfrac{1-\widetilde{n}}{\Delta\omega/\omega_{\mathrm{o}}}}$, in agreement with Eq.~\ref{Eq:SpectralTiltAngleSTWP} \cite{Hall21OL}.

We can thus interpret the bandwidth-dependence of $\delta_{\mathrm{o}}^{(1)}$ reported in \cite{Hall21OL} as a result of the non-differentiable frequency $\omega\!=\!\omega_{\mathrm{o}}$ representing a spectral barrier after which the spectrum cannot be extended. This occurs only in the close vicinity of $\omega_{\mathrm{o}}$ because of the rapid change in $\varphi_{\mathrm{c}}^{(1)}$ as the frequency $\omega_{\mathrm{c}}$ moves away from the non-differentiable point $\omega_{\mathrm{o}}$. When $\omega_{\mathrm{c}}$ is far from $\omega_{\mathrm{o}}$ so that the bandwidth does \textit{not} include $\omega_{\mathrm{o}}$ [Fig.~\ref{Fig:Parabola}(b)], $\tan{\delta_{\mathrm{o}}^{(1)}}$ is given by Eq.~\ref{Eq:SpectralTiltAngleSTWPAwayFromNDPoint} and the universality of Eq.~\ref{Eq:AngleOfPulseFrontTilt} is regained. A similar analysis applies to subluminal STWPs [Fig.~\ref{Fig:Parabola}(c,d)] where the non-differentiable frequency $\omega_{\mathrm{o}}$ is the maximum permissible frequency $\omega\!<\!\omega_{\mathrm{o}}$.

\section{Classification of pulsed fields after introducing AD}\label{Section:Classification}

\begin{figure*}[t!]
\centering
\includegraphics[width=16.5cm]{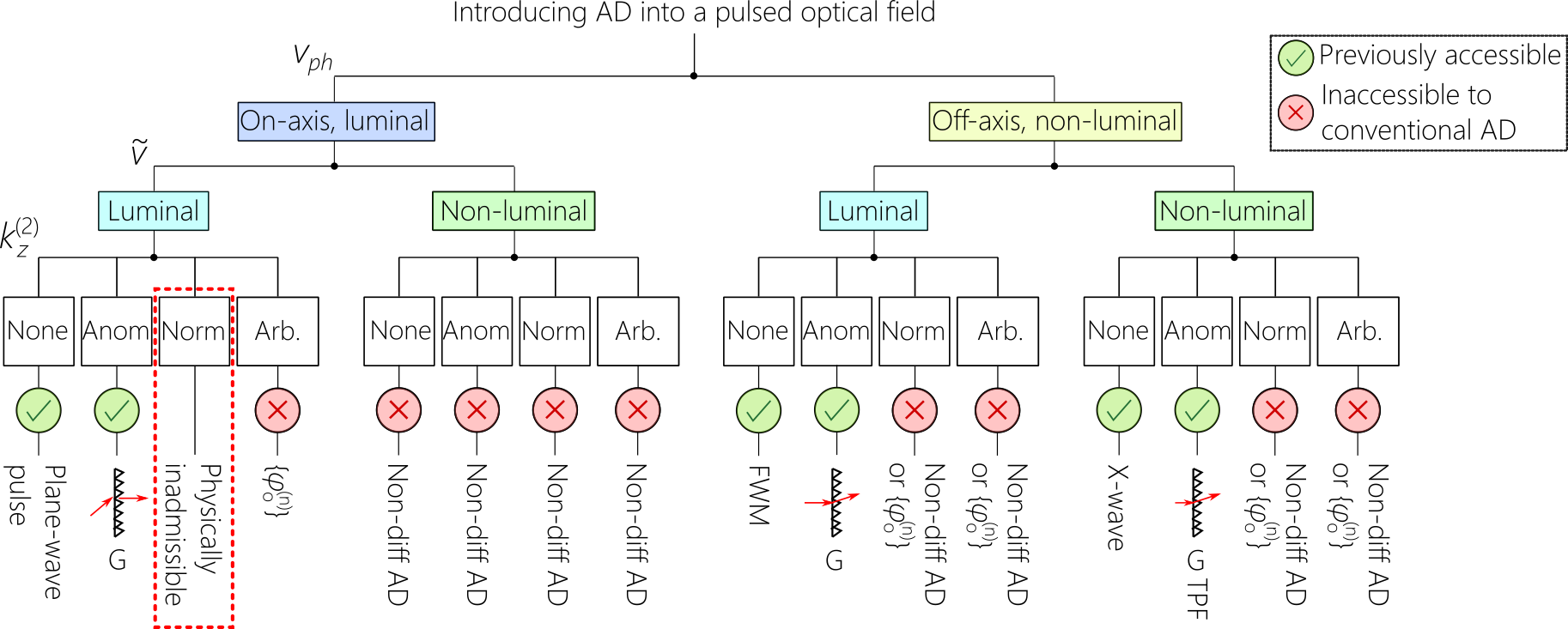}
\caption{Classification of pulsed optical fields incorporating AD. The first tier distinction is on-axis ($\varphi_{\mathrm{o}}\!=\!0$) propagation at luminal phase velocity $v_{\mathrm{ph}}=c$, or off-axis ($\varphi_{\mathrm{o}}\!\neq\!0$) propagation at non-luminal phase velocity $v_{\mathrm{ph}}\neq c$. The second tier is for group velocities that are luminal $\widetilde{v}=c$ or non-luminal $\widetilde{v}\neq c$. The third tier is based on the dispersion profile: the wave packet is either non-dispersive (`None'), is endowed with anomalous GVD (`Anom.'), normal GVD (`Norm.'), or an arbitrary dispersion profile (`Arb.'). Out of the possible 16 distinct classes of field configurations, one class is physically inadmissible, and 6 classes have been previously synthesized with conventional AD-engineering approaches. The remaining 9 classes require a universal AD synthesizer, either to inculcate non-differentiable AD (`non-diff.~AD'), or differentiable AD (`diff.~AD') involving independent control over multiple AD orders. The on-axis fields with non-luminal $\widetilde{v}$ (four classes) can only be produced with non-differentiable AD.}
\label{Fig:Classification}
\end{figure*}

We have shown in Section~\ref{Section:ADControl} that independent control over the first three AD coefficients $\varphi_{\mathrm{o}}$, $\varphi_{\mathrm{o}}^{(1)}$, and $\varphi_{\mathrm{o}}^{(2)}$ yields full control over three key characteristics of the field: the phase velocity $v_{\mathrm{ph}}$, group velocity $\widetilde{v}$, and GVD coefficient $k_{z}^{(2)}$ in free space (with diminishing higher-order dispersion terms). Because these parameters are useful in practice, we will utilize them to classify optical fields endowed with AD.

We classify the first two of these characteristics ($v_{\mathrm{ph}}$ and $\widetilde{v}$) as dichotomous. The phase velocity $v_{\mathrm{ph}}$ can be (1) luminal $v_{\mathrm{ph}}\!=\!c$, corresponding to on-axis propagation ($\varphi_{\mathrm{o}}\!=\!0$); or (2) non-luminal $v_{\mathrm{ph}}\!\neq\!c$, corresponding to off-axis propagation ($\varphi_{\mathrm{o}}\!\neq\!0$). We categorize the group velocity as (1) luminal $\widetilde{v}\!=\!c$; or (2) non-luminal $\widetilde{v}\!\neq\!c$, encompassing subluminal $\widetilde{v}\!<\!c$, superluminal $\widetilde{v}\!>\!c$, and negative-$\widetilde{v}$ wave packets. The third characteristic (GVD) can be classified according to the following scheme: (1) GVD-free, whereby all the dispersion coefficients along the observation axis are eliminated; (2) normal GVD $k_{z}^{(2)}\!>\!0$; (3) anomalous GVD $k_{z}^{(2)}\!<\!0$; and (4) arbitrary GVD (control over multiple dispersion orders). 

This classification scheme, depicted in Fig.~\ref{Fig:Classification}, is organized as a decision tree in three tiers. First, $\varphi_{\mathrm{o}}$ determines an initial branching into luminal ($\varphi_{\mathrm{o}}\!=\!0$) or non-luminal ($\varphi_{\mathrm{o}}\!\neq\!0$) phase velocity. Next, the bifurcation in the second tier into luminal or non-luminal group velocity involves both $\varphi_{\mathrm{o}}$ and $\varphi_{\mathrm{o}}^{(1)}$. Finally, branching in the third tier is according to the four classes of dispersion whose realization depends on $\varphi_{\mathrm{o}}$, $\varphi_{\mathrm{o}}^{(1)}$, and $\varphi_{\mathrm{o}}^{(2)}$.

This scheme therefore arranges optical fields endowed with AD into 16 classes, out of which only 6 have been realized with conventional procedures for AD-engineering, including the trivial case of the AD-free plane-wave pulse \cite{Hall24JOSAA}, while 10 classes are inaccessible. This classification surprisingly reveals that the majority of field configurations made possible by tuning the first AD orders have \textit{not been previously synthesized}. Part of the reason for this surprising gap is common misunderstandings regarding physically inadmissible fields; for example, (1) it is usually understood from the MGF theorem \cite{Martinez84JOSAA} that AD can\textit{not} produce normal GVD in free space, which thus eliminates 4 classes from contention; and (2) the standard perturbative theory of AD provides a compelling argument for dismissing on-axis wave packets with $\widetilde{v}\neq c$, which eliminates 3 more classes. Furthermore, because of the difficulty of controlling higher-order AD coefficients, 3 more fields are eliminated, corresponding to fields endowed with arbitrary dispersion profiles. We have shown that independent control over $\varphi_{\mathrm{o}}$, $\varphi_{\mathrm{o}}^{(1)}$, and $\varphi_{\mathrm{o}}^{(2)}$ is sufficient to produce normal GVD in free space in \textit{off-axis} fields, and that non-differentiable AD is necessary to produce normal GVD in \textit{on-axis} fields for tuning the group velocity in on-axis fields. This new understanding paves the way for new applications that harness the under-utilized potential of AD in optics.

One class is indeed physically inadmissible: on-axis luminal wave packets with normal GVD, whose spectral support necessarily lies below the light-line, thereby corresponding to purely evanescent waves. The remaining 9 classes can be synthesized using a universal AD synthesizer \cite{Hall24JOSAA} (see next Section). Of these 9 classes, 4 can be produced exclusively with non-differentiable AD, whereas the remaining 5 can be produced either with non-differentiable AD or with differentiable AD as long as independent control can be exercised over the first three AD coefficients.

\section{Universal angular-dispersion synthesizer}\label{Section:UniversalADSynthesizer}

\subsection{Why do we need a universal AD synthesizer?}

Conventional diffractive or dispersive optical devices are characterized by a small number of accessible physical degrees-of-freedom (whether structural or material); e.g., the ruling density of a grating, the chromatic dispersion in a prism, or the density and shape of meta-atom features in a metasurface. Consequently such devices control only the lowest two AD orders $\varphi_{\mathrm{o}}$ and $\varphi_{\mathrm{o}}^{(1)}$, which enables tuning $v_{\mathrm{ph}}$ and $\widetilde{v}$ in off-axis fields. No known optical device provides independent control even over $\varphi_{\mathrm{o}}^{(2)}$, which is needed to produce normal GVD in free space in an off-axis field, and no devices produce the non-differentiable AD necessary for synthesizing STWPs. For all of these reasons, it is useful to develop an optical system that is sufficiently versatile so as to synthesize the full scope of optical fields classified in Fig.~\ref{Fig:Classification}.

\subsection{What is a universal AD synthesizer?}

A universal AD synthesizer is an optical arrangement that endows a generic optical field (whether coherent or incoherent) with an arbitrary AD profile $\varphi(\omega)$. This corresponds to exercising independent control over the magnitude and sign of a large number of AD coefficients [Fig.~\ref{Fig:NonDiffAD}]. Ideally, such a universal AD synthesizer can be capable of introducing non-differentiable AD.

\subsection{Realization of a universal AD synthesizer}

\begin{figure}[t!]
\centering
\includegraphics[width=8.6cm]{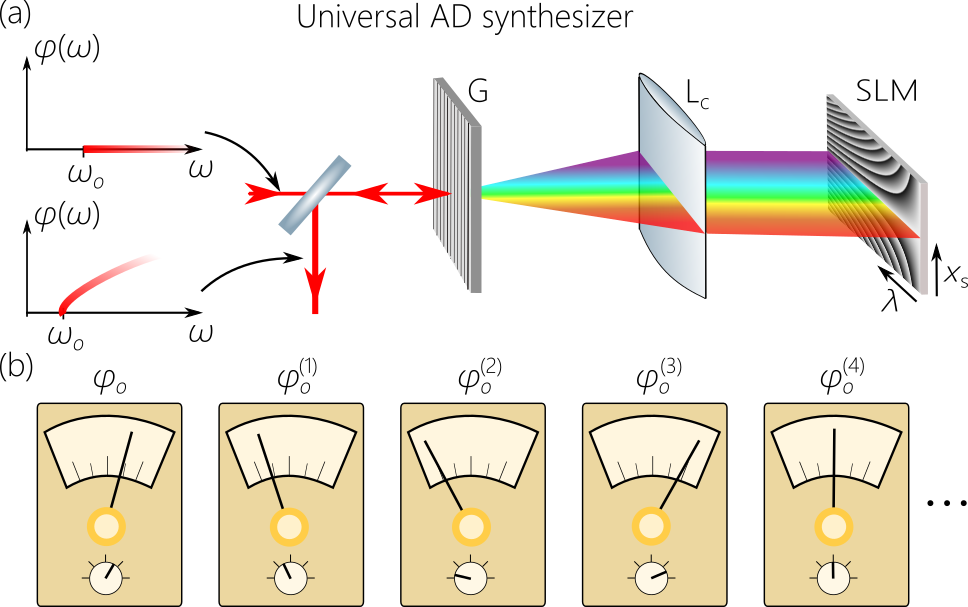}
\caption{(a) A universal AD synthesizer introduces an arbitrary angular profile $\varphi(\omega)$ in a plane-wave pulse. G: Diffraction grating; L$_{c}$: cylindrical lens; SLM: spatial light modulator. (b) All the AD coefficients in such a synthesizer are independently accessible, and are shown as `active' knobs, in contrast to the case of a grating in Fig.~\ref{Fig:DiffAD}.}
\label{Fig:NonDiffAD}
\end{figure}

We have introduced a universal AD synthesizer that makes use of the two-step methodology illustrated in Fig.~\ref{Fig:NonDiffAD}(a). First, the spectrum of an incident collimated broadband field is spatially resolved and collimated by a diffraction grating and a cylindrical lens. Each wavelength is now confined to a column in the focal plane. In this step we harness the exquisite spectral resolution afforded by high-quality gratings, but we do \textit{not} rely on the grating to introduce AD. In the second step, a device that modulates the phase of the spectrally resolved wave front, such as an SLM \cite{Kondakci17NP,Bhaduri18OE}, a phase plate \cite{Kondakci18OE,Bhaduri19OL}, or a metasurface \cite{Chen22SciAdv}, is placed at the focal plane to impart to each wavelength $\lambda$ a spatial phase distribution $\tfrac{2\pi}{\lambda}\sin{\{\varphi(\lambda)\}}x_{\mathrm{s}}$, where $x_{\mathrm{s}}$ is the coordinate along the SLM column and normal to the axis along which the spectrum is spread. This phase deflects the wavelength $\lambda$ at an angle $\varphi(\lambda)$ with the $z$-axis. The retro-reflected wave front traces its path back through the cylindrical lens to the grating, whereupon the wavelengths are recombined, and the AD-endowed field reconstituted. Alternatively, a transmissive phase plate or SLM can be utilized, and the modulated wave front is then transmitted to symmetrically placed cylindrical lens and grating in an unfolded configuration \cite{Kondakci18OE}.

We can change the phase slope associated with $\varphi(\lambda)$ at each column \textit{arbitrarily} without violating any physical principle, whether continuous or discontinuous, differentiable or non-differentiable. Therefore, this configuration is a \textit{universal} AD synthesizer capable of independently tuning a large number of AD coefficients [Fig.~\ref{Fig:NonDiffAD}(b)]. Another useful feature of this arrangement is that it can produce symmetrized fields (STWPs or TPFs) by assigning a pair of spatial frequencies $\pm\tfrac{2\pi}{\lambda}\sin{\{\varphi(\lambda)\}}x_{\mathrm{s}}$ to each wavelength $\lambda$. This can be done by splitting the $\omega$-column on the SLM into two halves along $x_{\mathrm{s}}$ centered on the propagation axis, and implementing each phase distribution in one of the halves. Finally, relative spectral \textit{phases} can be introduced between the frequencies just as in a conventional $4f$ pulse shaper \cite{Weiner00RSI}, and relative spectral \textit{amplitudes} can also be introduced via superpixel-encoding \cite{Tocnaye97AO}.

This configuration has produced:
\begin{enumerate}
\item STWPs that are propagation invariant over large distances \cite{Bhaduri18OE,Bhaduri19OL,Hall25OE1km}, with controllable group velocity \cite{Kondakci19NC,Bhaduri19Optica,Bhaduri20NP}, and axially encoded propagation characteristics, whether the on-axis wavelength (axial spectral encoding) \cite{Motz21PRA,Hall25OL} or the group velocity (axial acceleration) \cite{Yessenov20PRL2,Li20CP,Li20SR,Hall22OLAccel}.
\item STWPs incorporating controllable magnitude, sign, and order of GVD in free space \cite{Yessenov21ACSP}.
\item Discontinuous AD to tune the group velocity in different spectral windows \cite{Yessenov20NC}.
\item Discretized spatiotemporal spectra to explore a variety of space-time Talbot effects \cite{Yessenov20PRL1,Hall21APLP,Hall21OLTalbot}.
\item Hybrid guided spatiotemporal modes coupled to planar waveguides \cite{Shiri20NC,Shiri22ACSP}, multimode waveguides \cite{Bejot21ACSP,Shiri2022Optica,Shiri23JOSAA} and fibers \cite{Kibler21PRL,Stefanska23ACSP,Su25NC}, pulses that are omni-resonant with planar cavities \cite{Shiri20OL}, and that couple light to space-time surface plasmon polaritons \cite{Schepler20ACSP,Ichiji23PRA,Ichiji23ACSP,Ichiji24JOSAA,Ichiji25STSPP}.
\end{enumerate}

It is useful at this point to compare our strategy to the conventional approach for introducing AD via a grating. A grating adds a \textit{fixed} transverse wave number $m\tfrac{2\pi}{\Lambda}$ to \textit{all} the incident wavelengths, whereas the SLM provides a \textit{different} transverse wave number $\tfrac{2\pi}{\lambda}\sin{\{\varphi(\lambda)\}}$ to each wavelength, which effectively corresponds to implementing a multiplicity of distinct gratings -- one for each $\lambda$. Consequently, our strategy can provide the large number of independent parameters required to tune the angular dispersion coefficients $\{\varphi_{\mathrm{o}}^{(n)}\}$ and thus produce the desired dispersion profile $\{k_{z}^{(n)}\}$. Alternatively, this arrangement may be viewed as a grating whose \textit{period} is wavelength-dependent; that is, $\Lambda(\lambda)\!=\!\tfrac{m\lambda}{\sin{\varphi(\lambda)}}$. At $\varphi(\lambda)\!=\!0$, $\Lambda$ is infinite and the phase pattern is a constant; as $\varphi(\lambda)$ increases, $\Lambda(\lambda)$ decreases. Although the period implementable by an SLM is significantly larger than that in high-quality gratings, this is nevertheless compensated by the rapid rate of change in $\Lambda$ with $\lambda$ possible with an SLM.

\section{Synthesizing conical angular dispersion}
\label{2D_NDAD}

\begin{figure}[t!]
\centering
\includegraphics[width=8.6cm]{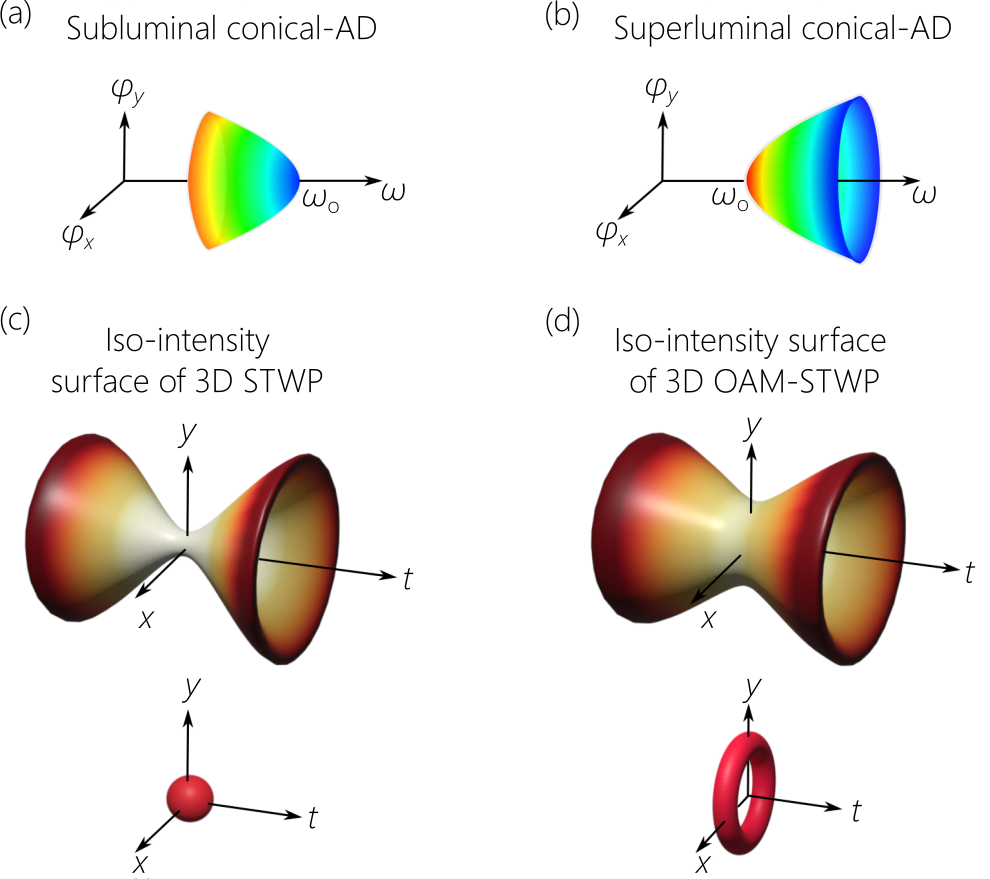}
\caption{(a) The spectral surface associated with conical-AD for a subluminal and (b) a superluminal STWP; $\omega_{\mathrm{o}}$ is the non-differentiable frequency. (c) Spatiotemporal intensity profile for an OAM-free STWP incorporating conical-AD. The lower panel is an iso-intensity surface at 0.9 the peak intensity. (d) Same as (c) after endowing the STWP with OAM number $\ell\!=\!1$. }
\label{Fig:2DAngularDisp}
\end{figure}

TPFs and the early developed propagation-invariant STWPs incorporate AD along one transverse dimension. Producing rotationally symmetric fields requires controlling AD over both transverse dimensions or conical-AD [Fig.~\ref{Fig:2DAngularDisp}(a,b)]. Such an extension can also facilitate studying orbital angular momentum (OAM) in the context of STWPs. Achieving this goal requires developing a universal conical-AD synthesizer.

The AD structure required to produce rotationally symmetric STWPs inculcating conical-AD is shown in Fig.~\ref{Fig:2DAngularDisp}(a) for the subluminal case [compare to Fig.~\ref{Fig:LightConesForSTWavePackets}(a)] and in Fig.~\ref{Fig:2DAngularDisp}(b) for the superluminal [compare to Fig.~\ref{Fig:LightConesForSTWavePackets}(b)]. The non-differentiable frequency $\omega_{\mathrm{o}}$ is now the spectral terminus of a 2D \textit{surface} rather than a 1D \textit{curve}. The associated spatiotemporal intensity structure is illustrated in Fig.~\ref{Fig:2DAngularDisp}(c) where we plot a 3D iso-intensity surface of the STWP, which takes on in general the same form for the subluminal and superluminal cases if they have similar spatial and temporal bandwidths. The X-shaped intensity profile has now been rotated cylindrically around $t$ (or, equivalently, the $z$-axis). By endowing this 3D STWP with OAM, a null is formed through the center of the spatiotemporal field structure [Fig.~\ref{Fig:2DAngularDisp}(d)].

To control conical-AD, the setup shown in Fig.~\ref{Fig:NonDiffAD}(c) is not useful because one dimension of the SLM is reserved for wavelength, so the field can be modulated spatially along only one dimension. This is indeed a fundamental restriction shared by all other approaches that spatiotemporally modulate an optical field with an SLM (e.g., spatiotemporal optical vortices \cite{Jhajj16PRX,Hancock19Optica,Chong20NP}). In all such cases, the temporal DoF is coupled to one spatial dimension of the field, leaving the second transverse spatial dimension separable. Circular gratings can be used to introduce conical-AD, but the conical-AD profile produced is differentiable and offers only limited control \cite{Piccardo23NP}.

A different approach has been developed for inculcating conical-AD \cite{Yessenov22NC}, which starts with spatially resolving the spectrum using chirped Bragg gratings, before implementing a 1D transformation along the direction of separated wavelengths to reshuffle the spectrum and arrange the wavelengths in a prescribed order. Next, a conformal geometric transformation, known as the log-polar transformation \cite{Bryngdahl74JOSA,Hossack87JOMO,Berkhout10PRL,Lavery12OE,Li19OE}, is implemented on the spectrally resolved wave front to map lines at its input to circles at its output. Therefore, the wavelengths that are initially ordered as lines in a prescribed sequence, are now curled into a continuous sequence of concentric circles. A spherical lens finally converts this spectrum into an STWP. Because the wavelengths can be organized radially in almost any sequence, this system is a universal conical-AD synthesizer. Much work is needed to realize with it the full gamut of STWPs that were previously produced with the universal AD synthesizer shown in Fig.~\ref{Fig:NonDiffAD}(a). To date we have produced propagation-invariant STWPs incorporating conical-AD with tunable group velocities that are localized in all dimensions \cite{Yessenov22NC} [Fig.~\ref{Fig:2DAngularDisp}(c)], the first propagation-invariant STWPs endowed with OAM \cite{Yessenov22NC} [Fig.~\ref{Fig:2DAngularDisp}(d)], STWPs endowed with polarization vortex structures \cite{Yessenov22OL}, and topological spin-texture spin textures imprinted on open and closed spectral surfaces \cite{Yessenov25Meron}.

A simplification is made possible when the source has a discrete spectrum (e.g., a laser comb). By resolving the spectrum, each spectral line can be spatially modulated independently. As mentioned above, this has been used in realizing novel Talbot effects in space-time \cite{Yessenov20PRL1,Hall21APLP,Hall21OLTalbot}. Moreover, each discrete wavelength can be associated with different OAM content \cite{Zou22OL,Pang22OL,Su24OE}, although such related optical fields do not -- strictly speaking -- incorporate AD.

\section{Non-differentiable angular dispersion as a resource}\label{Section:NDAD_as_a_resource}

In any realistic setting, it is physically impossible to produce an ideal AD profile $\varphi(\omega)$ in which each propagation angle is associated strictly with a single frequency, which requires infinite energy \cite{Sezginer85JAP}. This corresponds to a grating with infinite area in the universal AD synthesizer in Fig.~\ref{Fig:NonDiffAD}(a). In practice, there is always some finite spectral uncertainty $\delta\omega$ in the association between propagation angle $\varphi$ and frequency $\omega$. How `close' can a realistic field configuration approach ideal non-differentiable AD? 

We have recently introduced a `Schmidt number' that quantifies the non-differentiable AD as a resource \cite{Hall22JOSAA}. We write the envelope $\psi(x,z;t)$ from Eq.~\ref{Eq:PlaneWavePulse} in the more general form:
\begin{equation}
\psi(x,z;t)=\iint\!dk_{x}d\Omega\widetilde{\psi}(k_{x},\Omega)e^{i\{k_{x}x+(k_{z}-k_{\mathrm{o}})z-\Omega t\}},
\end{equation}
where the spatiotemporal spectrum $\widetilde{\psi}(k_{x},\Omega)$ is the Fourier transform of $\psi(x,0;t)$. The Schmidt decomposition \cite{Schmidt06MA}, a theoretical tool that has been useful in analyzing entangled photon states in quantum optics \cite{Ekert95AJP,Law00PRL,Law04PRL,Eberly06LP,Borshchevskaia19LPL}, can be defined for $\widetilde{\psi}(k_{x},\Omega)$ as follows:
\begin{equation}
\widetilde{\psi}(k_{x},\Omega)=\sum_{m}\sqrt{a_{m}}\chi_{m}(k_{x})\eta_{m}(\Omega),
\end{equation}
where $\{\chi_{m}(k_{x})\}$ and $\{\eta_{m}(\Omega)\}$ are two sets of orthonormal functions, a single index $m$ runs over both sets, and the Schmidt coefficients $\{a_{m}\}$ are normalized such that $\sum_{m}a_{m}\!=\!\iint\!dxdt|\psi(x,z;t)|^{2}\!=\!1$ for all $z$. This decomposition corresponds to the well-known singular-value decomposition in matrix algebra.

We define the Schmidt number $N_{\mathrm{S}}\!=\!\tfrac{1}{\sum_{m}a_{m}^{2}}$, which can be interpreted as the effective number of spatiotemporal mode pairs (one for $k_{x}$ and the other for $\Omega$) required to construct $\widetilde{\psi}(k_{x},\Omega)$. In the ideal limit of no spectral uncertainty $\widetilde{\psi}(k_{x},\Omega)\rightarrow\!\widetilde{\psi}(\Omega)\delta(k_{x}-k_{x}(\Omega))$, then $N_{\mathrm{S}}\!\rightarrow\!\infty$. Alternatively, when the spatiotemporal spectrum is separable $\widetilde{\psi}(k_{x},\Omega)\!\rightarrow\!\widetilde{\psi}_{x}(k_{x})\widetilde{\psi}_{t}(\Omega)$, then $N_{\mathrm{S}}\!\rightarrow\!1$. For finite spectral uncertainty, we have $N_{\mathrm{S}}\!>\!1$. We have shown that $N_{\mathrm{S}}$ drops when $\delta\omega$ increases \cite{Hall22JOSAA}. This indicates that reducing the spectral uncertainty requires control over a large number of independent degrees-of-freedom in the AD synthesizer. The parameter $N_{\mathrm{S}}$ can thus be viewed as a quantifier of this number.

\section{Roadmap for future developments}\label{Section:Roadmap}

The emerging concept of non-differentiable AD described here can be developed in several directions.

\textbf{(1) Role of polarization.} Only scalar fields were examined here, and only a few studies involving polarization have been reported \cite{Diouf21OE,Yessenov22OL}. Because conical-AD corresponds to a spectral surface, spin texture can be imbued by modulating the polarization at each point on this spectral surface, which enables investigations of topological optical structures in three dimensions \cite{Guo21Light,Yessenov25Meron}, such as skyrmions and merons \cite{Shen2023NP}. A universal vectorial conical-AD synthesizer is required for this application.

\textbf{(2) Omni-resonance.} Because the resonant linewidth of a planar Fabry-P{\'e}rot (FP) cavity is inversely proportional to the resonant field enhancement, there has been a longstanding effort in optical physics to broaden the resonant linewidth without impacting the cavity $Q$-factor. One such approach examined recently is `omni-resonance', whereby AD is introduced into the incident field to match the AD intrinsic to a single longitudinal mode of the cavity \cite{Shabahang17SR,Shiri20OL}. For oblique incidence on the cavity (off-axis field), differentiable AD suffices, but with a magnitude that is larger than that produced by a diffraction grating \cite{Shabahang17SR,Shabahang19OL}. For normal incidence, non-differentiable AD is required to achieve omni-resonance \cite{Shiri20OL,Shiri20APLP}. Such field configurations can harness the enhanced linear and nonlinear light-matter interactions benefiting from field buildup within the cavity over broad continuous spectra, rather than over narrow linewidths at discrete cavity resonances, including solar energy \cite{Villinger21AOM}. For such applications, a compact, large-area, alignment-free AD synthesizer is necessary.

\textbf{(3) Non-differentiable AD and light-matter interactions.} Only the surface has been scratched in terms of the use of non-differentiable AD in interactions of light with matter and photonic devices. It is expected that investigating the impact of non-differentiable AD on nonlinear and quantum optics will be significant. For example, phase-matching conditions induce AD, but with large spectral uncertainty, so that a universal AD synthesizer can help engineer these phase-matching conditions, an area that awaits broader investigations.

\textbf{(4) Coherent versus incoherent optical fields.} We have dealt exclusively here with coherent optical pulses. However, the same analysis can be extended to incoherent fields, such as produced by a super-luminescent diode (SLD) or a spatially filtered LED. Initial results for incoherent STWPs incorporating non-differentiable AD include demonstrating diffraction-free propagation \cite{Yessenov19Optica}, propagation invariance of the spatiotemporal coherence function \cite{Yessenov19Optica}, and tunable coherence velocity (the group velocity of the coherence function) \cite{Yessenov19OL}. Such an extension can benefit from establishing a coherence theory for spatiotemporally structured optical fields.

\textbf{(5) New avenues for AD synthesizers.} A crucial question is whether a single device such as a metasurface or other nano-structured devices, rather than bulk free-space optics, can inculcate non-differentiable AD into an optical field or independently modify multiple AD orders. Can lasers (whether bulk or on-chip) be modified to directly produce light endowed with non-differentiable AD; that is, an STWP laser? Recent strides have been taken to reduce the size of the universal AD synthesizer in Fig.~\ref{Fig:NonDiffAD}(a) \cite{Yessenov23OL} using a new class of chirped Bragg gratings \cite{Mhibik23OL,Mhibik23OL2}. Much remains to be studied, especially in the realm of conical-AD synthesizers. It is clear that synergy with nanophotonics is crucial to push this enterprise forwards.

\textbf{(6) Other applications.} STWPs have been used in a host of other applications, such as optical delay lines \cite{Yessenov20NC}, self-healing \cite{Kondakci18OL}, and a host of anomalous refractive phenomena at planar interfaces \cite{Bhaduri20NP,Motz21OL,Yessenon21JOSAA1,Motz21JOSAA,Yessenon21JOSAA2}. These characteristics are particularly useful for biomedical imaging, an area that is just emerging \cite{Diouf22SA} and which has yet to significantly benefit from spatiotemporally coupled fields. Finally, fundamental studies of spatiotemporally structured fields as an extension to continuous degrees-of-freedom of the optical field of the concept of classical entanglement await \cite{Kagalwala13NP,Forbes19PO}.

\section{Conclusions}\label{Section:Conclusions}

In conclusion, we have provided a perspective on two new concepts related to the well-established phenomenon of AD in optics: (1) \textit{non-differentiable AD}, which helps circumvent a host of restrictions on the behavior of optical fields; and (2) \textit{universal AD synthesizers}, which introduce arbitrary AD profiles in an optical field, and are thus capable of introducing non-differentiable AD or controlling multiple AD orders independently. We have focused on the specific novel aspects characterizing non-differentiable AD that contrast with its conventional differentiable counterpart produced via traditional optical components.

These newly emerging fundamental concepts in optical physics stand to have profound impact on a broad range of applications, which are now being explored, extending from solar energy and biomedical imaging to nonlinear and quantum optics. Synergy with nanophotonics in particular will be crucial to reduce the size of universal AD synthesizers, which will make non-differentiable AD accessible in compact devices.

\section*{Acknowledgments}

This work was funded by the U.S. Office of Naval Research under contract N00014-20-1-2789.

\bibliography{diffraction}

\end{document}